\documentclass[11pt,a4paper]{article}
\usepackage{jheppub}
\usepackage{epsf}
\usepackage{amsmath,amssymb,amsfonts}
\usepackage{bbold}
\usepackage{graphicx,graphics}
\usepackage{epstopdf}
\usepackage{url}
\usepackage{relsize,exscale,scalefnt,anyfontsize}

\newcommand{\nn}{\nonumber}

\newcommand{\ie}{{\em i.e.}}
\newcommand{\eq}[1]{(\ref{#1})}
\newcommand{\bvec}{{\boldsymbol b}}
\newcommand{\svec}{{\boldsymbol s}}
\newcommand{\Ep}{E_\mathrm{p}}
\newcommand{\mprot}{m_\mathrm{p}}
\newcommand{\mT}{M_\perp}
\newcommand{\jpsi}{{\mathrm J}/\psi}
\newcommand{\xf}{x_{\mathrm{F}}}
\newcommand{\pt}{p_{_\perp}}
\newcommand{\dd}{{\rm d}}
\newcommand{\half}{\frac{1}{2}}  
\newcommand{\lsim}{\lesssim} 
\newcommand{\qzero}{\hat{q}_0}
\newcommand{\qzeropm}{\hat{q}_0^\pm}
\newcommand{\gevsqfm}{GeV$^2$/fm}
\newcommand{\sigmann}{\sigma_{_{\rm NN}}}
\newcommand{\Leff}{L_{\rm eff}}
\newcommand{\LeffA}{L_{\rm eff}^{\rm A}}
\newcommand{\LeffB}{L_{\rm eff}^{\rm B}}
\newcommand{\ndf}{{\rm ndf}}
\newcommand{\be}{\begin{equation}}
\newcommand{\ee}{\end{equation}}
\newcommand{\bea}{\begin{eqnarray}}
\newcommand{\eea}{\end{eqnarray}}
\newcommand{\ella}{L_{_{\rm A}}}
\newcommand{\ellb}{L_{_{\rm B}}}
\newcommand{\qhat}{\hat{q}}
\newcommand{\rab}{R_{\rm AB}}
\newcommand{\rcucu}{R_{\rm CuCu}}
\newcommand{\rauau}{R_{\rm AuAu}}
\newcommand{\rpbpb}{R_{\rm PbPb}}

\newcommand{\rpa}{R_{\rm pA}}
\newcommand{\rpb}{R_{\rm pB}}
\newcommand{\dya}{{\delta y_{_{\rm A}}}}
\newcommand{\dyb}{{\delta y_{_{\rm B}}}}
\newcommand{\dymax}{{\delta y^{\rm max}}}
\newcommand{\Eb}{E_{_{\rm B}}}
\newcommand{\Phat}{\hat{{\cal P}}}
\newcommand{\LeffAmean}{\langle \Leff^{\rm A} \rangle}

\newcommand{\LeffAC}{\langle \Leff^{\rm A} \rangle_{_{\cal C}}}

\newcommand{\Npart}{\langle N_{\rm part} \rangle}
\newcommand{\NpartC}{\langle N_{\rm part} \rangle_{_{\cal C}}}
\newcommand{\epsb}{\varepsilon_{_{\rm B}}}
\newcommand{\epsbmax}{\varepsilon^{\rm max}}
\newcommand{\TA}{T_{\rm A}}
\newcommand{\TB}{T_{\rm B}}
\newcommand{\TAB}{T_{\rm AB}}
\newcommand{\sqrts}{\sqrt{s}}
\def\pt{p_{_\perp}}

\title{Quarkonium suppression in heavy-ion collisions \\ from coherent energy loss in cold nuclear matter }

\author[a,b]{Fran\c{c}ois Arleo,}
\author[c]{St\'ephane Peign\'e}

\affiliation[a]{Laboratoire Leprince-Ringuet (LLR), \'Ecole polytechnique, CNRS/IN2P3 91128 Palaiseau, France}
\affiliation[b]{Laboratoire d'Annecy-le-Vieux de Physique Th\'eorique (LAPTh)\\ UMR5108, Universit\'e de Savoie, CNRS, BP 110, 74941 Annecy-le-Vieux cedex, France}
\affiliation[c]{SUBATECH, UMR 6457, Universit\'e de Nantes, Ecole des Mines de Nantes, IN2P3/CNRS \\ 4 rue Alfred Kastler, 44307 Nantes cedex 3, France}

\emailAdd{francois.arleo@cern.ch}
\emailAdd{peigne@subatech.in2p3.fr}

\abstract{
The effect of parton energy loss in cold nuclear matter on the suppression of quarkonia ($\jpsi$, $\Upsilon$) in heavy-ion collisions is investigated, by extrapolating a model based on coherent radiative energy loss recently shown to describe successfully $\jpsi$ and $\Upsilon$ suppression in proton--nucleus collisions. Model predictions in heavy-ion collisions at RHIC (Au--Au, Cu--Cu, and Cu--Au) and LHC (Pb--Pb) show a sizable suppression arising from the sole effect of energy loss in cold matter. This effect should thus be considered in order to get a reliable baseline for cold nuclear matter effects in quarkonium suppression in heavy-ion collisions, in view of disentangling hot from cold nuclear effects. 
}


\begin{document}

\maketitle

\section{Introduction and summary}

Quarkonium production in heavy-ion (A--A) collisions is a widely discussed observable, expected to be highly sensitive to the presence of a hot medium. However, predicting quarkonium production rates in A--A collisions is a difficult task, due to various competing effects, such as quarkonium suppression from Debye screening in a hot medium~\cite{Matsui:1986dk} and enhancement due to recombination processes at high energy~\cite{BraunMunzinger:2000px,Thews:2000rj}. In order to interpret reliably the heavy-ion measurements, accurate \emph{baseline} predictions assuming only cold nuclear effects are needed.

Until quite recently, most of the phenomenological approaches assumed either nuclear absorption, nuclear parton distribution function (nPDF) or saturation effects to be responsible for $\jpsi$ and $\Upsilon$ suppression in proton--nucleus (p--A) collisions. In a series of papers, however, we argued that medium-induced parton energy loss in cold nuclear matter could play a decisive role in the suppression of $\jpsi$ and $\Upsilon$ states (denoted as $\psi$ in the following) in p--A collisions~\cite{Arleo:2010rb,Arleo:2012hn,Arleo:2012rs,Arleo:2013zua}. Quite remarkably, all available $\psi$ suppression measurements from fixed-target experiments (SPS, HERA, FNAL) to RHIC could be described within a simple model and on a broad kinematical range in rapidity~\cite{Arleo:2012rs} and transverse momentum~\cite{Arleo:2013zua}. What is more, predictions in p--Pb collisions at the LHC proved in excellent agreement with ALICE~\cite{Abelev:2013yxa} and LHCb~\cite{Aaij:2013zxa} data.

We emphasize that this successful description of the $\psi$ suppression data at various collision energies 
can be obtained, with a very good $\chi^2/{\ndf}$, without including nPDF effects~\cite{Arleo:2012rs}. On the contrary, nPDF/shadowing effects alone cannot achieve such a global description, in particular they fail to describe the shape of $\psi$ suppression as a function of $\xf$ (or of the rapidity $y$),  
at both fixed-target and collider energies.\footnote{The difficulty for nPDF effects to produce a correct parametric dependence of $\psi$ suppression is well-known: if shadowing effects would play a dominant role for $\psi$ nuclear suppression, the latter should scale in the target momentum fraction, $x_2$. However a drastic violation of $x_2$-scaling is observed when comparing the $\psi$ suppression data at various collision energies.} 
This suggests switching the way to apprehend cold nuclear effects
 in quarkonium suppression, namely, to consider parton energy loss as the leading effect. Shadowing/nPDF effects may affect the {\it magnitude} of $\psi$ suppression\footnote{The effect of nPDFs on the magnitude of $\psi$ suppression in p--A collisions is minor at fixed-target, and sizable at collider energies~\cite{Arleo:2012rs}.} but not so much the shape (at least in $\xf$ or $y$), and might thus be viewed as `corrections' which do not change the qualitative picture of $\psi$ suppression obtained with parton energy loss alone. Let us also mention that medium-induced parton energy loss is as fundamental as nPDFs, and could actually apply more generally. As a matter of fact, it should also play a role in processes breaking 
QCD factorization, for which a description in terms of nPDFs alone would be unfounded.

The agreement of the model used in Refs.~\cite{Arleo:2012hn,Arleo:2012rs,Arleo:2013zua} with $\psi$ suppression data in p--A collisions originates mainly from the parametric behavior $\Delta E \propto E$ (where $E$ is the $\psi$ energy) of the medium-induced radiative parton energy loss. In particular this behavior is essential to describe the increase of $\psi$ suppression with increasing rapidity. This parametric law arises when a fast incoming color charge crosses the target nucleus and is scattered to small angle (in the target rest frame). It is thus expected to hold in quarkonium hadroproduction, where typically a high-energy gluon from the projectile proton is scattered to a compact color octet heavy $Q \bar{Q}$ pair~\cite{Arleo:2010rb}. The radiative loss $\Delta E \propto E$ originates from gluon radiation which is {\it fully coherent} over the size $L$ of the nucleus. The parametric dependence (in $E$, $L$ and the mass $M$ of the produced compact color state) of the coherent radiation spectrum and associated average loss $\Delta E$ was first derived in \cite{Arleo:2010rb}, and recently reviewed in \cite{Peigne:2014uha} in a fully defined theoretical setup. Note that the same coherent, medium-induced radiation spectrum arises in the production of a single forward particle \cite{Arleo:2010rb,Peigne:2014uha} and in forward dijet production \cite{Liou:2014rha,Peigne:2014rka}, suggesting the broad relevance of coherent energy loss in hard forward p--A processes.

In the present study we extrapolate the model of Refs.~\cite{Arleo:2012hn,Arleo:2012rs,Arleo:2013zua} to nucleus--nucleus collisions. To illustrate the main idea, consider the production of a compact color octet $Q \bar{Q}$ pair through gluon-gluon fusion, at mid-rapidity (and low transverse momentum, $p_\perp \lsim M$) 
in the nucleon--nucleon c.m. frame of some A--B collision. In the rest frame of the nucleus B, the $Q \bar{Q}$ pair is produced at large (positive) rapidity from the incoming fast gluon of the `projectile' A. This leads to a suppression of the $\psi$ production rate due to coherent energy loss induced by rescatterings in B, as shown in~\cite{Arleo:2012rs} in the case of p--B collisions. Analogously, when viewed in the rest frame of A, $\psi$ production at large (negative) rapidity must be affected by the coherent energy loss induced by rescatterings in A, leading to an additional $\psi$ suppression at $y=0$ in the c.m. frame.\footnote{We expect the two effects to add incoherently, because rescatterings in A and B induce gluon radiation spectra which populate different regions of phase space, see section~\ref{sec:aa} and Fig.~\ref{fig-rap-regions}.} 

The goal of the present study is to provide baseline predictions, based on coherent energy loss through {\it cold} nuclear matter, for the rapidity and centrality dependence of $\jpsi$ and $\Upsilon$ suppression in heavy-ion collisions at RHIC and LHC. The model is first generalized to deal with nucleus--nucleus collisions in section~\ref{sec:model}. The predictions at RHIC and LHC are given in sections~\ref{sec:rhic} and~\ref{sec:lhc}, where they are also compared to the heavy-ion data. In the final discussion, section~\ref{sec:discussion}, we argue that the observed discrepancies between the baseline predictions and the data are qualitatively consistent with the presence of additional {\it hot} suppression effects in heavy-ion collisions, and of recombination processes in the specific case of $\jpsi$ production at the LHC. Although the present study focusses 
on parton energy loss, we also shortly discuss nPDF effects in section~\ref{sec:discussion}. Using different nPDF sets, we roughly 
estimate $\jpsi$ suppression due to nPDF effects alone, and observe that it never exceeds the strength of the suppression assuming only coherent energy loss. In particular, at the LHC and at large enough rapidity, the effect of energy loss on $\jpsi$ suppression dominates over the effect of nPDFs.

\section{Model for quarkonium suppression in nuclear collisions}
\label{sec:model}

\subsection{Proton--nucleus collisions}
\label{sec:pa}

We briefly remind in this section the basics of the model based on coherent energy loss used to describe $\psi$ suppression measured in proton--nucleus collisions. The single differential p--B production cross section as a function of the $\psi$ energy reads~\cite{Arleo:2012rs}
\be
\label{eq:xspB0-energy}
\frac{1}{B}\frac{\dd\sigma_{\mathrm{pB}}^{\psi}}{\dd \Eb} \left( \Eb \right)  = \int_0^{\epsbmax} \dd \epsb \,{\cal P}(\epsb, \Eb, \ell_{_{\rm B}}^2) \, \frac{\dd\sigma_{\mathrm{pp}}^{\psi}}{\dd \Eb} \left( \Eb+\epsb \right) \, ,
\ee
where $\Eb$ (respectively, $\epsb$) is the energy (respectively, energy loss) of the $Q \bar{Q}$ pair in the rest frame of the nucleus B. The upper limit on the energy loss is $\epsbmax=\min\left(\Eb,\Ep-\Eb\right)$, where $\Ep$ is the beam energy in that frame, and the p--p production cross section is given by a fit to data. ${\cal P}$ denotes the energy loss probability distribution, or \emph{quenching weight}.

The quenching weight is related to the medium-induced, coherent radiation spectrum $\dd I/\dd\varepsilon$ given in~\cite{Arleo:2012rs} (and earlier in \cite{Arleo:2010rb}), which is a very good approximation to the exact spectrum computed to all orders in the opacity expansion~\cite{Peigne:2014uha}. For convenience the explicit expression of ${\cal P}$ is quoted in Appendix \ref{app-quenching}. It depends on the accumulated transverse momentum transfer $\ell_{_{\rm B}} = \sqrt{\qhat \ellb}$ (assumed to satisfy  $\ell_{_{\rm B}} \ll \mT$) due to soft rescatterings in nucleus B, where $\ellb$ is the path-length discussed in section~\ref{sec:length} and $\qhat$ the transport coefficient in cold nuclear matter. More precisely~\cite{Arleo:2012rs},
\be
\label{qhat-model}
\hat{q} \equiv \hat{q}_0 \left[ \frac{10^{-2}}{\min(x_0, x_2)} \right]^{0.3}\ ; \ \ \  x_0 \equiv \frac{1}{2 m_\mathrm{p} \ellb}\ ; \ \ \ x_2 \equiv  \frac{\mT}{\sqrts} \, e^{-y} \, ,
\ee
where $y$ is the $\psi$ rapidity in the center-of-mass frame of the proton--nucleon collision (of energy $\sqrts \simeq \sqrt{2 \mprot \Ep}$, with $\mprot$ the proton mass), and $M_\perp=(M^2+\pt^2)^{\half}$ is the transverse mass of the $Q \bar{Q}$ pair. In the present paper, we consider quarkonium production integrated over $\pt$ (and thus dominated by typical values $\pt < M$) for which using $2 \to 1$ kinematics for the partonic subprocess is a reasonable simplifying assumption. 
The value of $\qzero$ used in this analysis and its uncertainty are discussed in section~\ref{sec:uncertainties}. For clarity, the dependence of $\qhat$ on the medium size $L$ (through the value of $x_0$, see \eq{qhat-model}) will be implicit in the following.

In view of generalizing the model to A--B collisions in the next section, where the projectile and target play symmetric roles, it is convenient to change variable from $\Eb$ to the (proton--nucleon) c.m. frame rapidity $y$, using 
\be\label{eq:Ey}
\Eb= \Ep \, \frac{\mT}{\sqrts} \, e^{y} \equiv E(y) \, .
\ee
From \eq{eq:xspB0-energy} we obtain
\be
\label{eq:xspB0}
\frac{1}{B}\frac{\dd\sigma_{\mathrm{pB}}^{\psi}}{\dd y} = \int_0^{\epsbmax(y)} \dd\epsb \,
{\cal P}(\epsb, E(y), \qhat(y)\ellb) \,\left[ \frac{E(y)}{E(y)+\epsb} \right] \, \frac{\dd\sigma_{\mathrm{pp}}^{\psi}}{\dd y} \left( y\left(E(y)+\epsb\right) \right) \, ,
\ee
where $\epsbmax(y)=\min\left(E(y),\Ep-E(y)\right)$. We now change the integration variable and express the energy loss $\epsb$ in terms of a shift in rapidity, $\dyb$, defined as 
\be
E(y)+\epsb  \equiv E(y+\dyb) = E(y) \, e^{\dyb} \ \ \Leftrightarrow \ \  \dyb = \ln{\left( 1 + \frac{\epsb}{E(y)}  \right)} \, .
\ee
Using the fact that the quenching weight is a scaling function of the variable $x= \varepsilon / E$, namely $E\,{\cal P}(\varepsilon,E,\ell^2) = \Phat(x \equiv \varepsilon/E,\ell^2)$, 
we can rewrite \eq{eq:xspB0} as
\be
\label{eq:xspB}
\frac{1}{B}\frac{\dd\sigma_{\mathrm{pB}}^{\psi}}{\dd y}\left(y \right) = 
\int_0^{\dymax(y)}  \dd\dyb \,
\Phat(e^\dyb-1, \qhat(y)\ellb) \, \, \frac{\dd\sigma_{\mathrm{pp}}^{\psi}}{\dd{y}} \left( y+\dyb \right) \, .
\ee
Here $\dymax(y)= \min\left(\ln{2},y_{\mathrm{max}}-y\right)$, with $y_{\mathrm{max}} = \ln(\sqrts/\mT)$ the maximal $\psi$ rapidity (in the proton--nucleon c.m. frame) allowed by our kinematics.\footnote{For $\jpsi$ production, taking $M=3\,{\rm GeV}$ for the $c \bar{c}$ pair and $\pt=1\,{\rm GeV}$, we obtain $y_{\mathrm{max}} \simeq 4.1$ at RHIC ($\sqrts = 200\,{\rm GeV}$) and $y_{\mathrm{max}} \simeq 6.8$ at LHC ($\sqrts = 2.76\,{\rm TeV}$).}

The expression \eq{eq:xspB}, together with the explicit form of $\Phat$ given in Appendix \ref{app-quenching} (see \eq{quenching-dilog}), was used in \cite{Arleo:2012rs,Arleo:2013zua} to study $\psi$ nuclear suppression in proton-nucleus collisions.

\subsection{Nucleus--nucleus collisions}
\label{sec:aa}

Let us now consider the more complicated case of quarkonium production in nucleus--nucleus collisions. In a generic A--B collision both incoming partons, respectively from the `projectile' nucleus A and the `target' nucleus B, might suffer multiple scattering in the nucleus B and A, respectively. Consequently, gluon radiation off both partons can interfere with that of the final state particle (here, the compact color octet $Q\bar{Q}$ pair), making a priori difficult the calculation of the medium-induced gluon spectrum in the collision of two heavy ions.

However, consider gluon radiation induced by rescattering in nucleus B. The medium-induced, coherent gluon spectrum is found in \cite{Arleo:2010rb,Peigne:2014uha} to arise from the logarithmic $k_\perp$ domain
\be
\label{kperp-range}
\frac{\omega}{\Eb} \, M_\perp \ll k_\perp \ll \ell_{_{\rm B}} \, ,
\ee
where $\omega$ and $k_\perp$ denote the radiated gluon energy (in the B rest frame) and transverse momentum, and $\ell_{_{\rm B}} \ll \mT$ is the typical transverse broadening in nucleus B already introduced in section~\ref{sec:pa}.\footnote{\samepage Obviously the logarithmic range \eq{kperp-range} only appears when $\omega \ll \Eb \ell_{_{\rm B}}/M_\perp$, \ie, when $\omega$ is much smaller than the typical $\omega$ contributing to the average energy loss $\Delta E$~\cite{Arleo:2010rb}. In the present study we focus on such $\omega$ values, which can be checked to dominate in the convolution \eq{eq:xspB0-energy} (where $\epsb = \omega$), due to the fast decrease of the p--p cross section with increasing rapidity~\cite{Arleo:2012rs}.}

\begin{figure}[ht]
\centering
\includegraphics[width=11cm]{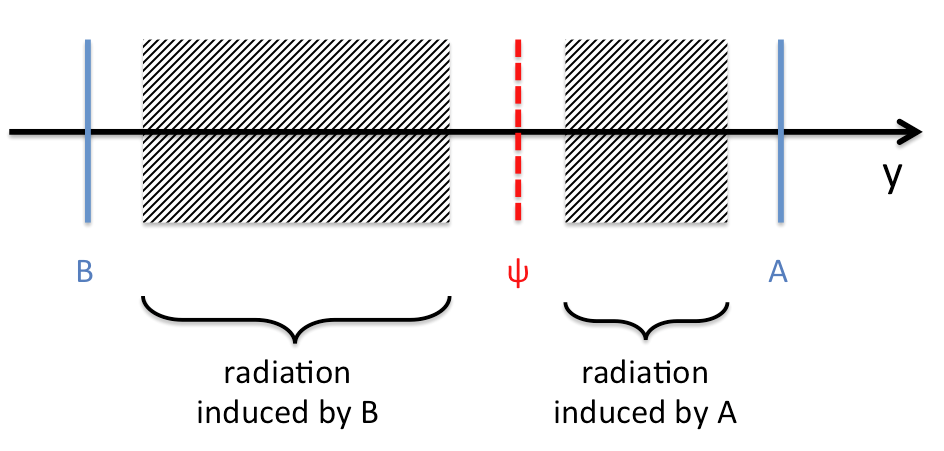}
\caption{Sketch of the rapidity regions populated by medium-induced radiation in an A--B collision. The `target' B and `projectile' A move with respectively negative and positive rapidities.}
\label{fig-rap-regions}
\end{figure}

Introducing the rapidities of the $\psi$ state and of the radiated gluon in the B frame,  
\be
y_{\rm B}^\psi = \frac{1}{2} \ln{ \left(\frac{\Eb + p_{_{\rm B}}^z}{\Eb - p_{_{\rm B}}^z}\right)} \simeq \ln{ \left(\frac{2\Eb}{M_\perp}\right)} \ \ ; \ \ y_{\rm B}^g= \frac{1}{2} \ln{ \left(\frac{\omega + k^z}{\omega - k^z}\right)} \simeq \ln{ \left(\frac{2 \omega}{k_\perp}\right)} \, ,
\ee
the leftmost inequality in \eq{kperp-range} becomes
\be
\frac{\omega}{k_\perp} \ll \frac{\Eb}{M_\perp} \ \Rightarrow \  \exp{(y_{\rm B}^g)} \ll \exp{(y_{\rm B}^\psi)} \ \Rightarrow \ y_{\rm B}^g < y_{\rm B}^\psi  \, .
\ee
The latter inequality must hold in all longitudinally boosted frames, namely, $y^g < y^\psi$. Thus, the medium-induced radiation associated to rescattering in B populates the region of rapidities {\it smaller} than the $\psi$ rapidity. Similarly, the induced radiation associated to rescattering in A populates $y^g > y^\psi$. This is illustrated in Fig.~\ref{fig-rap-regions}. 

Since gluon radiation induced by rescattering in nuclei A and B occurs in distinct regions of phase space, 
it can be combined in a probabilistic manner as follows. We first 
express the $\psi$ production cross section in A--B simply as a function of that in A--p collisions using \eq{eq:xspB}
\be
\label{eq:xsAA}
\frac{1}{A B}\frac{\dd\sigma_{\mathrm{AB}}^{\psi}}{\dd y}\left(y\right)  = \int_{_0}^{^{\dymax(y)}} \hskip -8mm \dd\dyb \,\Phat(e^{\dyb}-1, \qhat(y)\ellb) \, \, \frac{1}{A}\frac{\dd\sigma_{\mathrm{Ap}}^{\psi}}{\dd{y}} \left( y+\dyb\right) \, .
\ee
Then, using again \eq{eq:xspB} to express the A--p cross section as a function of that in p--p collisions, one obtains\footnote{Since rescattering processes happen simultaneously in A and B, the result \eq{eq:xsAA2} should not depend on the order in which the energy losses induced by A and B are taken into account. We therefore neglect the shift $\dyb$ in $\hat{q}$ and $\dymax$ when using \eq{eq:xspB} to go from \eq{eq:xsAA} to \eq{eq:xsAA2}. Namely, $\hat{q}(-y-\dyb)\simeq\hat{q}(-y)$ and $\dymax(-y-\dyb)\simeq\dymax(-y)$. This ensures that \eq{eq:xsAA2} verifies $\dd\sigma_{_{\mathrm{BA}}}(y)/\dd{y}=\dd\sigma_{_{\mathrm{AB}}}(-y)/\dd{y}$.}
\be
\label{eq:xsAA2}
\frac{1}{AB}\frac{\dd\sigma_{\mathrm{AB}}^{\psi}}{\dd y}\left(y\right)= \int_{_0}^{^{\dymax(y)}} \hskip -1cm \dd\dyb \,\Phat(x_{_{\rm B}}, \qhat(y)\ellb) \int_{_0}^{^{\dymax(-y)}} \hskip -1.2cm \dd\dya \,\Phat(x_{_{\rm A}}, \qhat(-y) \ella) \, \frac{\dd\sigma_{\mathrm{pp}}^{\psi}}{\dd{y}} \left(y+\dyb-\dya\right) \, ,
\ee
where $x_{_{\rm A}} \equiv e^{\dya}-1$, $x_{_{\rm B}} \equiv e^{\dyb}-1$, and we used the fact that $\dd\sigma_{\mathrm{pp}}^{\psi}/{\dd{y}}$ is an even function of the rapidity. From \eq{eq:xsAA2} we can compute the nuclear suppression factor in (minimum bias) heavy-ion collisions,
\be
\label{eq:raa}
R_{\rm AB}\left(y\right) = \frac{1}{A B}\frac{\dd\sigma_{\mathrm{AB}}^{\psi}}{\dd y}\left(y\right) \biggm/ \frac{\dd\sigma_{\mathrm{pp}}^{\psi}}{\dd y}\left(y\right) \, .
\ee
Let us note that \eq{eq:xsAA2} can be rewritten as
\be
\label{eq:xsAA3}
\frac{1}{AB}\frac{\dd\sigma_{\mathrm{AB}}^{\psi}}{\dd y}\left(y\right)= \int \dd \delta y \,  \Phat_{\rm AB}(\delta y) \, \frac{\dd\sigma_{\mathrm{pp}}^{\psi}}{\dd{y}} \left(y+\delta y \right) \, ,
\ee
where $\Phat_{\rm AB}(\delta y)$ is the energy `loss' ($\delta y> 0$) or `gain' ($\delta y < 0$) probability distribution in A--B collisions, 
\be
\label{Phat-AB}
\Phat_{\rm AB}(\delta y) \equiv \int_{_0}^{^{\dymax(y)}} \hskip -1cm \dd\dyb \,\Phat(x_{_{\rm B}}, \qhat(y)\ellb ) \int_{_0}^{^{\dymax(-y)}} \hskip -1.2cm \dd\dya \,\Phat(x_{_{\rm A}}, \qhat(-y)\ella) \, \delta(\delta y-\dyb+\dya) \, .
\ee

\subsection{Medium length in the Glauber model}
\label{sec:length}

In minimum bias A--B collisions (\ie, after integration over the impact parameter $\bvec$), the effective path length covered by the compact color octet in nucleus A can be calculated in Glauber theory and shown to coincide with an expression derived for minimum bias p--A collisions~\cite{Arleo:2012rs}, 
\be
\label{Leff2}
\LeffAmean- L_{\rm p} =  \frac{(A-1)}{A^2\,\rho_0} \, \int \dd^2\svec \,  \TA^2(\svec) \, ,
\ee
and a similar expression for the effective path length in nucleus B. In Eq.~\eq{Leff2} the thickness function $\TA(\svec)$ is normalized as $\int \dd^2\svec \, \TA(\svec) = A$, and we use $L_{\rm p}=1.5$~fm for the length in a proton target and $\rho_0=0.17$~fm$^{-3}$ for the nuclear density, consistently with~\cite{Arleo:2012rs}.

In order to exhibit its centrality dependence, $\psi$ suppression is often measured as a function of the number of participants $\NpartC$ corresponding to a given centrality class ${\cal C}$, given by~\cite{Bialas:1976ed}
\bea
\label{Npart}
\NpartC =   \frac{1}{\sigma_{{\rm AB}}^{{\cal C}}}\ & \Bigg[ & \int_{\cal C} \dd^2\bvec \,\int \dd^2\svec \,  \TA(\svec)\ \left\{ 1 - {\left[ 1-  \frac{\sigmann}{B}\ \TB(\bvec - \svec) \right]}^B \right\} \nonumber \\ 
 &+& \int_{\cal C} \dd^2\bvec \,\int \dd^2\svec \, \TB(\svec)\ \left\{ 1 - {\left[ 1-  \frac{\sigmann}{A}\ \TA(\bvec - \svec) \right]}^A \right\} \Bigg] \ .
\eea
Here $\sigmann$ is the nucleon--nucleon inelastic cross section (we take $\sigmann=42$~mb at RHIC~\cite{Miller:2007ri} and $\sigmann=62.8$~mb~\cite{Abelev:2012sea} at LHC)
and $\sigma_{{\rm AB}}^{{\cal C}}$ is the A--B cross section of that centrality class, 
\be
\sigma_{{\rm AB}}^{{\cal C}} = \int_{\cal C} \dd^2\bvec \ \left\{ 1 - {\left[ 1-  \frac{\sigmann}{AB}\ \TAB(\bvec) \right]}^{AB} \right\} \, ,
\ee
where $\TAB(\bvec)\equiv \int d^2s\ \TA(\svec)\ \TB(\bvec-\svec)$.

In a given centrality class ${\cal C}$ of A--B collisions, the effective path length of the compact color octet across A can be estimated as
\be
\label{Leff2C}
\LeffAC- L_{\rm p} = \frac{(A-1)}{A \, \rho_0} \int_{\cal C} \dd^2\bvec \,\int \dd^2\svec \,  \TA^2(\svec)\ \TB(\bvec - \svec) \bigg/ \int_{\cal C} \dd^2\bvec \,  \TAB(\bvec) \, ,
\ee
and a similar expression for the path length across B. 
For the  0-100\% centrality class, \ie, integrating over all impact parameters in Eq.~\eq{Leff2C}, we recover the minimum bias expression \eq{Leff2}. 

In the following, we will use in~\eq{eq:xsAA2} the length \eq{Leff2C} to compute $\rab$ as a function of $\NpartC$, Eq.~\eq{Npart}, and the length \eq{Leff2} to compute $\rab$ as a function of $y$ in minimum bias collisions. 

\subsection{A simple approximation}
\label{sec:approx}

In order to get a baseline of cold nuclear matter effects expected in heavy-ion collisions, a data-driven extrapolation of proton--nucleus measurements has been used at RHIC~\cite{GranierdeCassagnac:2007aj,Adare:2007gn}. Assuming that final-state absorption and nPDF effects are the dominant cold nuclear effects, it is assumed in these studies that the suppression in A--B collisions is given by that in p--A and p--B collisions, according to
\be
\label{eq:approx_paap}
\rab(y) \simeq \rpa(-y) \times \rpb(+y) \, ,
\ee
where $\rpb$ is the cross section ratio,
\be
\label{eq:rpa}
\rpb\left(y\right) = \frac{1}{B}\frac{\dd\sigma_{\mathrm{pB}}^{\psi}}{\dd y}\left(y\right) \biggm/ \frac{\dd\sigma_{\mathrm{pp}}^{\psi}}{\dd y}\left(y\right) \, .
\ee

Here we show that this approximation also holds in the present energy loss model. Using \eq{eq:xspB} and  \eq{eq:xsAA2} we readily find 
\bea
\frac{1}{AB}\frac{\dd\sigma_{\mathrm{AB}}^{\psi}}{\dd y}\left(y \right)  = \int_0^{\dymax(-y)} \dd\dya \,\Phat(e^{\dya}-1, \qhat(-y) \ella) \, \rpb\left(y - \dya \right) \, \frac{\dd\sigma_{{\rm pp}}^{\psi}}{\dd y}\left(y - \dya \right) \, .  && \nn \\ 
&&
\eea
Now assuming that $\rpb(y)$ is a much smoother function than $\dd\sigma_{{\rm pp}}^{\psi}/{\dd y}$, one directly gets
\be
\label{eq:approx}
\frac{1}{AB}\frac{\dd\sigma_{\mathrm{AB}}^{\psi}}{\dd y}\left(y \right) \simeq \rpb\left(y \right)  \int_0^{\dymax(-y)} \dd\dya \,\Phat(e^{\dya}-1, \qhat(-y) \ella) \, \frac{\dd\sigma_{{\rm pp}}^{\psi}}{\dd y}\left(- y + \dya \right) \, . 
\ee
Using again \eq{eq:xspB}, and the definitions \eq{eq:raa} and \eq{eq:rpa}, one directly gets Eq.~\eq{eq:approx_paap}.

\begin{figure}[ht]
\centering
\includegraphics[width=8.5cm]{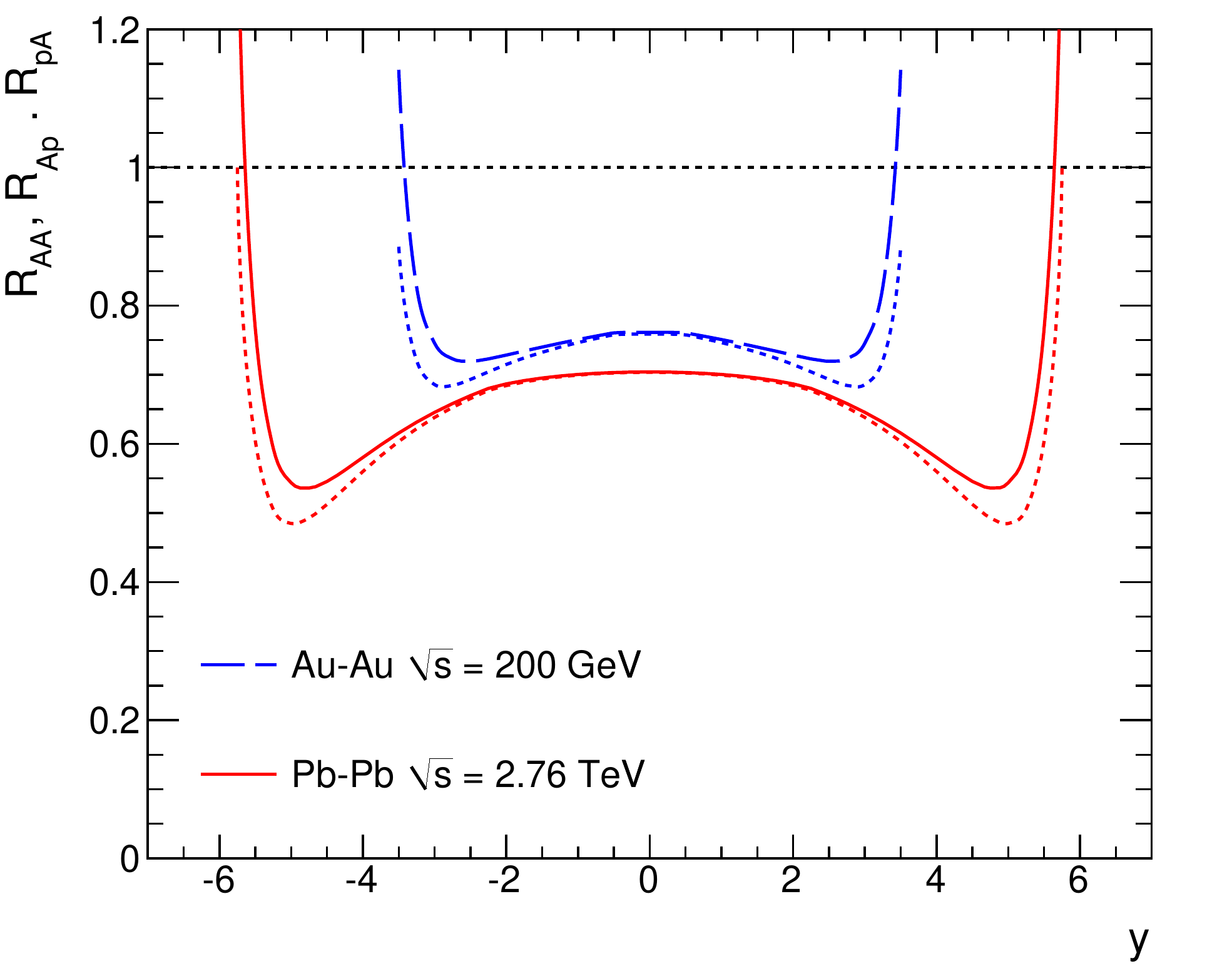}
\caption{Rapidity dependence of $\jpsi$ suppression from coherent energy loss, in minimum bias Au--Au collisions at RHIC (long-dashed line) and Pb--Pb collisions at LHC (solid) in comparison to the approximation \eq{eq:approx_paap} (dashed).}
\label{fig:approx_aa_paap}
\end{figure} 

In Fig.~\ref{fig:approx_aa_paap} we compute $\jpsi$ suppression in minimum bias 
Au--Au collisions at RHIC and Pb--Pb collisions at LHC using Eq.~\eq{eq:xsAA2} and 
the approximation Eq.~\eq{eq:approx_paap}. We used $\LeffAmean = 10.21$~fm for A$=$Au and $\LeffAmean = 10.11$~fm for A$=$Pb~\cite{Arleo:2012rs}. A smooth decrease of $\rab$ is observed as a function of $|y|$, until a value of the rapidity above which $\rab$ increases rapidly. The rise of $\rab$ at large rapidity is due to energy `gain' fluctuations, $\delta y < 0$ in \eq{eq:xsAA3}, which dramatically enhance the $\jpsi$ production cross section, $\sigma_{\mathrm{pp}}^{\psi} \left(y+\delta y \right) \gg \sigma_{\mathrm{pp}}^{\psi} \left(y \right)$ as $y$ is getting closer to the boundary of phase-space where $\sigma_{\mathrm{pp}}^{\psi} \left(y \right)$ becomes vanishingly small. Figure \ref{fig:approx_aa_paap} illustrates that the approximation~\eq{eq:approx_paap} proves remarkable.\footnote{Eq.~\eq{eq:approx_paap} remains quite accurate even when the variation of $\rpa$ with $y$ is fast and the approximation \eq{eq:approx} is in principle no longer justified.} Let us stress that the increase of $\rab$ at large $|y|$ seen on Fig.~\ref{fig:approx_aa_paap} appears outside the region of validity $|y| < |y^{\rm crit}|$ of our model (see section \ref{sec:validity}), and might be spoiled by nuclear absorption effects. However, note that the approximation~\eq{eq:approx_paap} is best precisely in the validity domain of our model. 

In order to estimate cold nuclear matter effects in Pb--Pb collisions at $\sqrts=2.76$~TeV, the ALICE collaboration determined the product~\cite{Abelev:2013yxa,Hadjidakis:2014tea}
\be
\label{eq:approx_paap_ALICE}
{R_{\rm Pb\,p}}(y_{\rm lab} = y-\Delta y, \sqrts=5\ {\rm TeV}) \times R_{{\rm pPb}}(y_{\rm lab} = y+\Delta y, \sqrts=5\ {\rm TeV}) \, , 
\ee
from their p--Pb measurements at $\sqrts=5$~TeV, where $\Delta y= 0.465$ (respectively $-\Delta y= -0.465$) is the boost of the center-of-mass with respect to the laboratory frame in p--Pb (respectively Pb--p) collisions. However, we find that the coherent energy loss effects do not follow exactly this extrapolation. In the acceptance of the ALICE muon spectrometer ($2.5 \leq |y| \leq 4$), differences range from 10 to 20\%. 

\subsection{Computing uncertainties}
\label{sec:uncertainties}

Equation~\eq{eq:xsAA2} is used to compute $\jpsi$ and $\Upsilon$ suppression in heavy-ion collisions. It requires the knowledge of the magnitude of the transport coefficient, $\qzero$, as well as the `slope' $n$ of the p--p cross section, parametrized as
$\dd\sigma_{\mathrm{pp}}^{\psi}/ \dd y  \propto \left(1- \frac{2 M_\perp}{\sqrt{s}} \cosh{y} \right)^{n}$~~\cite{Arleo:2012rs}.\footnote{Note that the normalization of the p--p cross section is irrelevant here as it cancels out when computing the 
factor $R_{\rm AB}$, Eq.~(\ref{eq:raa}).}

In order to assess the uncertainties of the model predictions, both quantities are varied around their central value, $S^0\equiv\{\qzero, n\}$. For a given quarkonium state ($\jpsi$ or $\Upsilon$) and at a given center-of-mass energy, four predictions are made assuming the following sets of parameters, $S_1^\pm=\{\qzeropm, n\}$ and $S_2^\pm=\{\qzero, n^\pm\}$, on top of the central prediction assuming $S^0$.
The (asymmetric) uncertainties of the model predictions are determined following the prescription suggested in~\cite{Pumplin:2001ct},
\begin{eqnarray}
\label{eq:errors}
\left(\Delta \rab^+\right)^2 & = & \sum_k \left[ \max\left\{ \rab(S^+_k)-\rab(S^0), \rab(S^-_k)-\rab(S^0),0 \right\} \right]^2 \ ,  \nonumber\\
\left(\Delta \rab^-\right)^2 & = & \sum_k \left[ \max\left\{ \rab(S^0)-\rab(S^+_k), \rab(S^0)-\rab(S^-_k),0 \right\} \right]^2 \ .
\end{eqnarray}

Let us now specify the estimated range for $\qzero$ and $n$. The transport coefficient $\qzero$ is the only free parameter of the model. It is determined by fitting the $\jpsi$ suppression measured by E866~\cite{Leitch:1999ea} in p--W over p--Be collisions ($\sqrt{s}=38.7$~GeV), 
see~\cite{Arleo:2012rs}. The obtained value is $\qzero=0.075\pm0.005$~\gevsqfm, with a slightly larger central value, $\qzero = 0.087$~\gevsqfm, when the fitting $\xf$-range is 
shrinked to a domain where the E866 data is the most precise. 
We thus use (in unit~\gevsqfm) $\{\qzero, \qzero^-, \qzero^+\}=\{0.075, 0.070, 0.09\}$. 

Regarding the slope of the p--p cross section, we shall use at RHIC the values obtained from the fit of the p--p data at $\sqrts=200$~GeV given in~\cite{Arleo:2012rs}, $\{n, n^-, n^+\}=\{8.3, 7.2, 9.4\}$ for $\jpsi$ 
and $\{n, n^-, n^+\}=\{6.7, 5.7, 7.7\}$ for $\Upsilon$. 
At the LHC, the p--p data at $\sqrts=7$~TeV give $n=32.3\pm 7.5$ for $\jpsi$ and $n=14.2\pm 2.9$ for $\Upsilon$. Lacking precise measurements in p--p collisions at $\sqrts=2.76$~TeV, we shall use the following slightly smaller, empirical values $\{n, n^-, n^+\}=\{22.5, 20, 25\}$ ($\jpsi$) and $\{n, n^-, n^+\}=\{12.5, 10, 15\}$ ($\Upsilon$) inferred from the $\sqrts$ dependence of $n$ by a power-law interpolation from RHIC to LHC.

\subsection{Range of validity}
\label{sec:validity}

The present picture is not expected to hold when the $\psi$ hadronizes in either the projectile nucleus A or target nucleus B and could thus suffer nuclear absorption. Denoting 
by $t_\psi(y)= (E(y)/M_\perp) \cdot \tau_\psi $ (where $\tau_\psi\simeq 0.3$~fm is the $\psi$ proper formation time) and $t_\psi(-y)$ the $\psi$ formation time in the rest frame of B and A, respectively, hadronization  occurs outside each nucleus when $t_\psi(y) \gtrsim \LeffB$ and $t_\psi(-y) \gtrsim \LeffA$. Using Eq.~(\ref{eq:Ey})
and $\Ep \simeq s/(2 \mprot)$, this condition translates into the following range of validity for the rapidity,
\be
\label{rap-range}
y^{\rm crit}(\sqrt{s}, B)  < y < - y^{\rm crit}(\sqrt{s}, A) \equiv \ln\left( \frac{\tau_\psi}{\LeffA} \cdot \frac{\sqrts}{2 \mprot} \right)  \, . 
\ee
Using $\LeffA \simeq 10$~fm for A$=$Au or A$=$Pb~\cite{Arleo:2012rs}, we obtain $y^{\rm crit} \simeq -1.2$ for Au--Au collisions at RHIC ($\sqrt{s} = 200\,{\rm GeV}$), and $y^{\rm crit} \simeq -3.8$ for Pb--Pb collisions at LHC ($\sqrt{s} = 2.76\,{\rm TeV}$). While at LHC the constraint \eq{rap-range} should be fulfilled, the RHIC measurements in the 
rapidity bins $|y| \simeq 2$ lie at the edge of the applicability of the model.

\section{RHIC}
\label{sec:rhic}

In this section the model predictions on quarkonium 
suppression in A--B collisions at RHIC are compared to PHENIX 
and STAR data, as a function of the rapidity 
(in a given centrality class) in section~\ref{sec:rhicrap}, and as a function of $\Npart$ (in a given rapidity bin) in section~\ref{sec:rhiccentrality}. The interpretation of the differences between the energy loss model predictions and the data is postponed to section~\ref{sec:discussion}.

\subsection{Rapidity dependence}
\label{sec:rhicrap}

\begin{figure}[h]
\centering
\includegraphics[width=5.1cm]{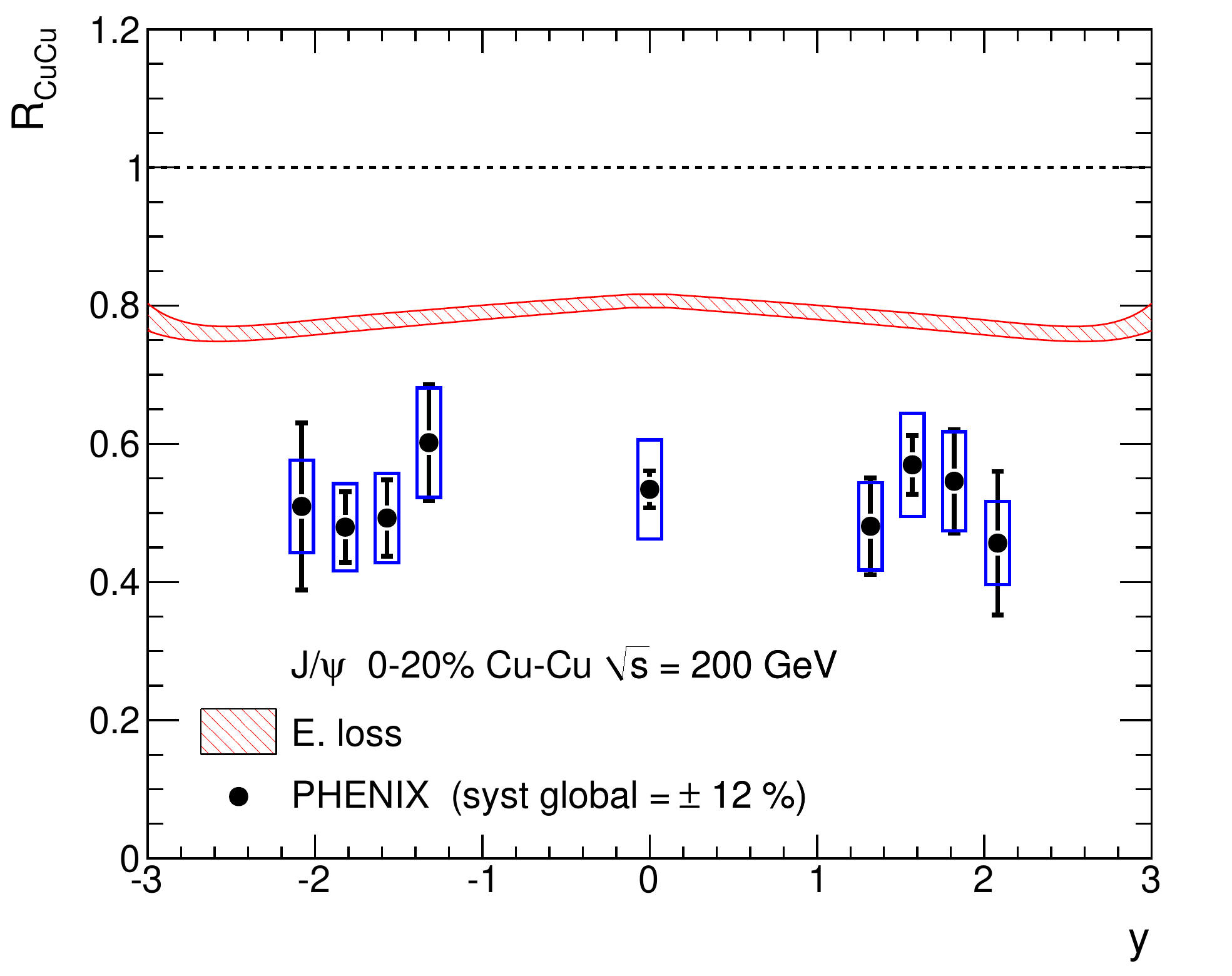} \hskip -3mm 
\includegraphics[width=5.1cm]{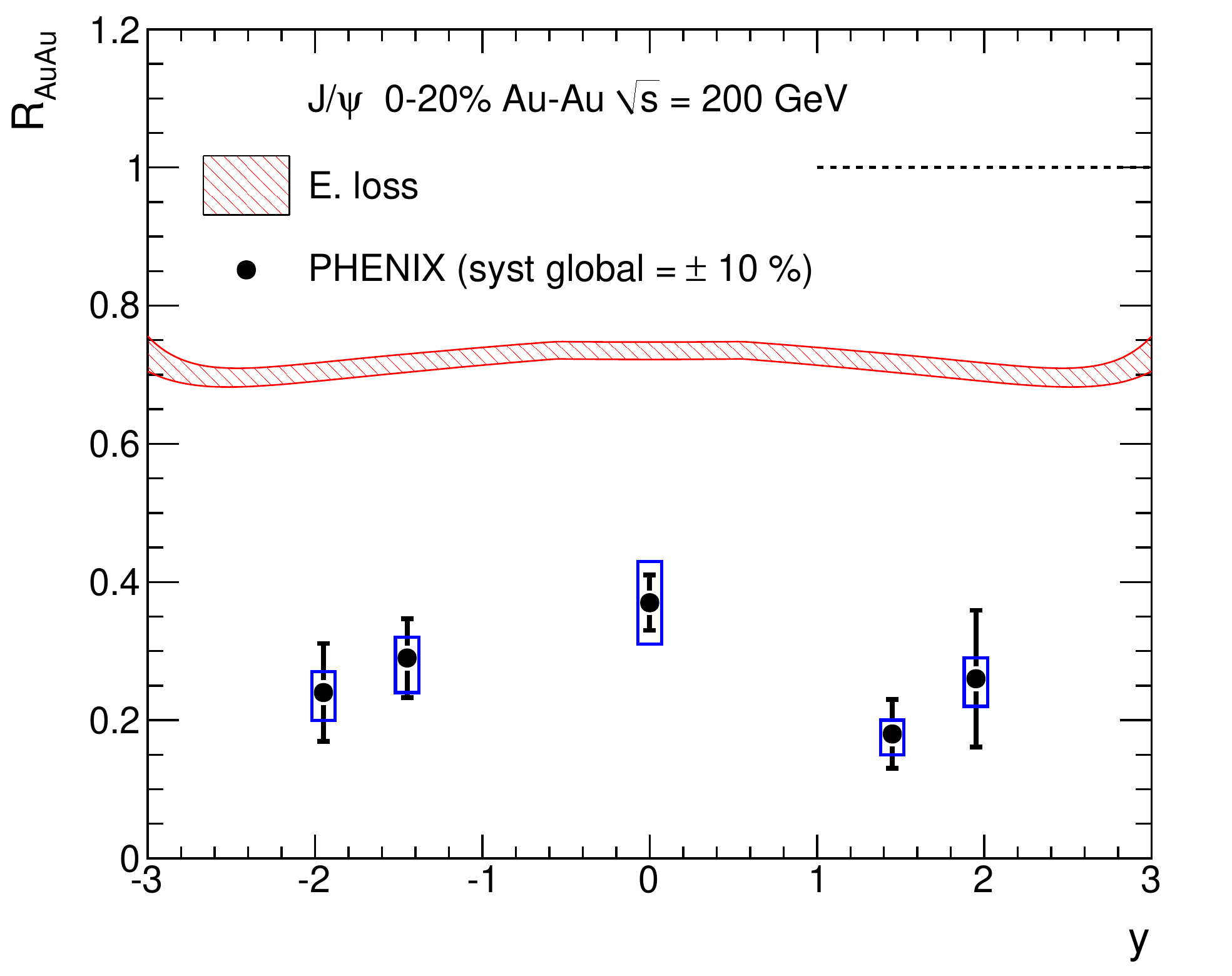} \hskip -3mm
\includegraphics[width=5.1cm]{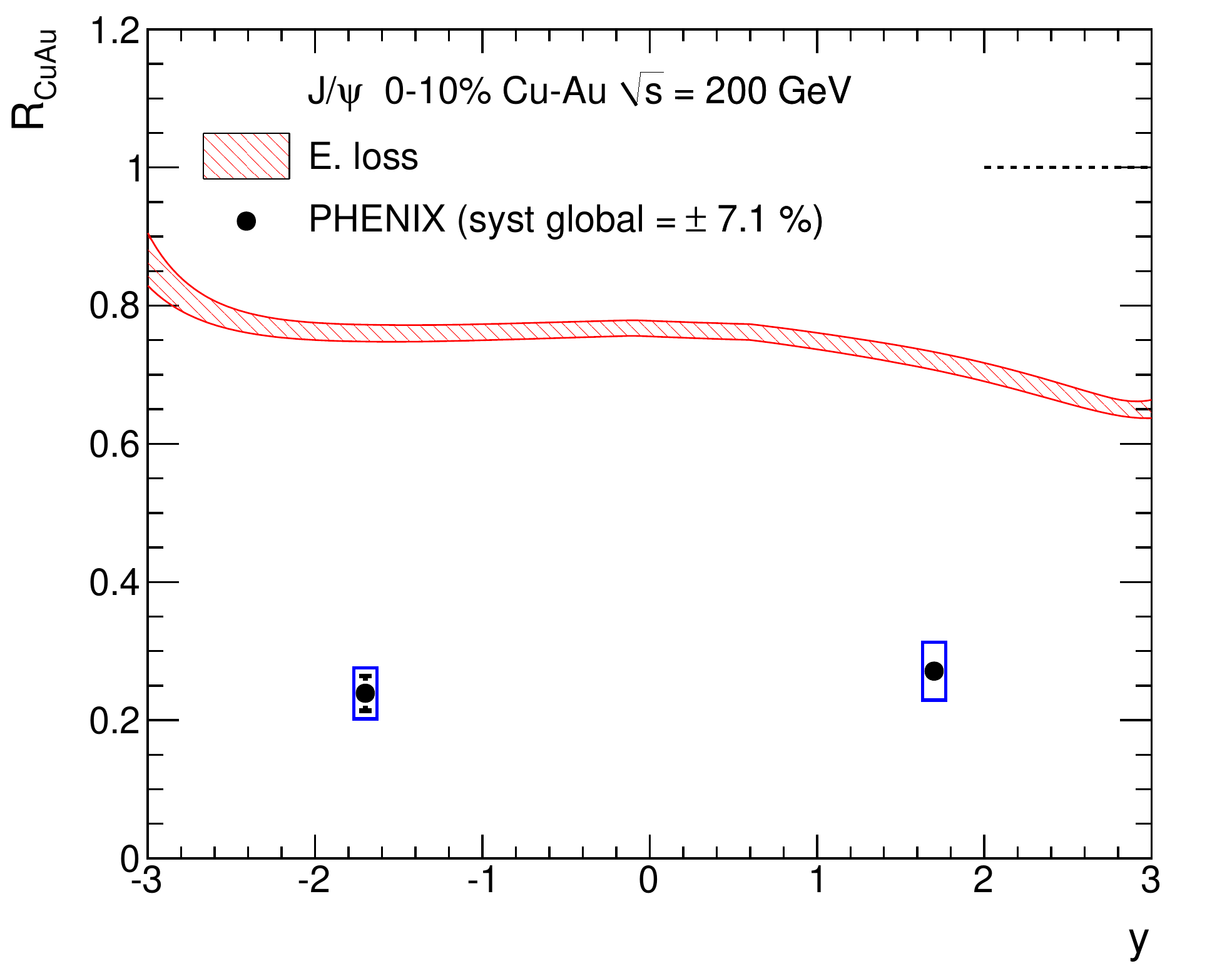}
\caption{Rapidity dependence of $\jpsi$ suppression in the $0$--$20\%$ most central Cu--Cu (left), Au-Au (middle), and   $0$--$10\%$ most central Cu--Au (right) collisions at $\sqrt{s}=200$~GeV predicted by the energy loss model (dashed band). PHENIX data are from~\cite{Adare:2008sh,Adare:2006ns,Aidala:2014bqx}.}
\label{fig:rhic_jps_cucu_auau}
\end{figure}

The rapidity dependence of $\jpsi$ suppression is computed in Fig.~\ref{fig:rhic_jps_cucu_auau} (dashed band) in central Cu--Cu (left), Au--Au (middle), and Cu--Au (right) collisions at $\sqrts=200$~GeV. The suppression is almost independent of the rapidity in the considered range ($|y|<3$), and slightly more pronounced in Au--Au collisions, $\rauau \simeq 0.75$ (at $y=0$), than in Cu--Cu collisions, $\rcucu \simeq 0.8$, because of the larger medium length encountered by the $c\bar{c}$ pair in the former collision system. In asymmetric (Cu--Au) collisions, $\rab$ is no longer an even function of $y$.
At negative rapidity, say $-2 < y < 0$, the suppression in p--Au (or, d--Au) collisions is rather moderate~\cite{Arleo:2012rs}, $R_{{\rm pAu}}(y) \lesssim 1$, and so is $R_{{\rm pCu}}(y) \simeq R_{{\rm pAu}}(y)$. Using~\eq{eq:approx_paap} one thus gets $R_{{\rm CuAu}}(y) \simeq R_{{\rm CuCu}}(y)$. At positive rapidity, however, the suppression due to coherent energy loss becomes more pronounced, and thus $R_{{\rm pAu}}(y) < R_{{\rm pCu}}(y) <1$, leading to a stronger suppression in Cu--Au with respect to Cu--Cu collisions.

The measurements by PHENIX~\cite{Adare:2008sh,Adare:2006ns,Aidala:2014bqx} are also shown for comparison.\footnote{In addition to the PHENIX data,  the STAR experiment also performed $\jpsi$ measurements in Au--Au collisions~\cite{Adamczyk:2013tvk}, yet with a larger uncertainty.} 
The suppression reported in Cu--Cu collisions is significantly more pronounced 
than the energy loss model prediction. 
In Au--Au collisions the discrepancy is even more striking: the suppression is $\rauau \sim 0.2$--$0.4$, \ie, 2 to 3 times smaller than the model expectations. Similarly, the PHENIX Cu--Au data exhibit a much stronger suppression than that predicted from the sole effect of energy loss. See section~\ref{sec:discussion} for a discussion. 

\subsection{Centrality dependence}
\label{sec:rhiccentrality}

\begin{figure}[hb]
\centering
\includegraphics[width=7.5cm]{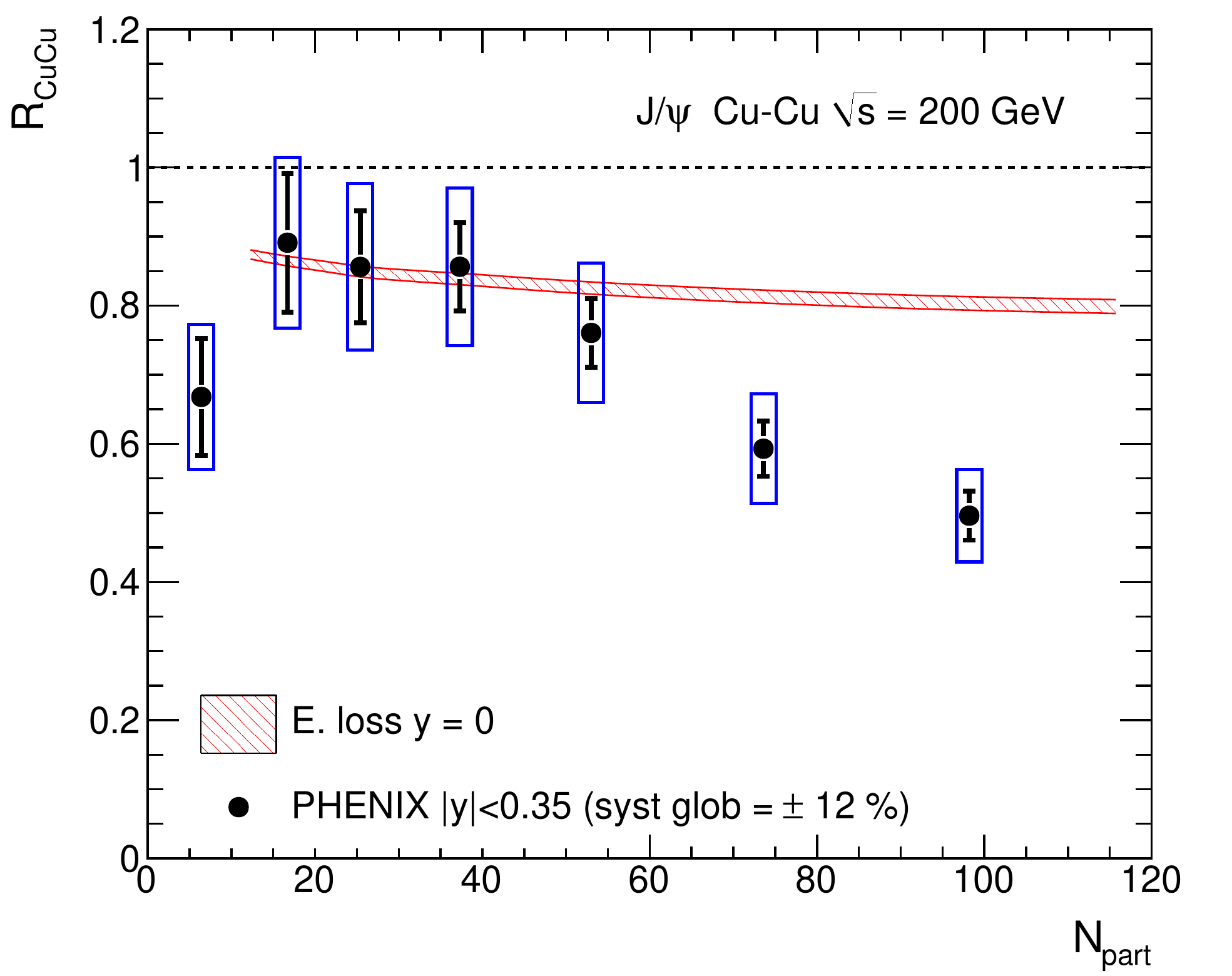}
\includegraphics[width=7.5cm]{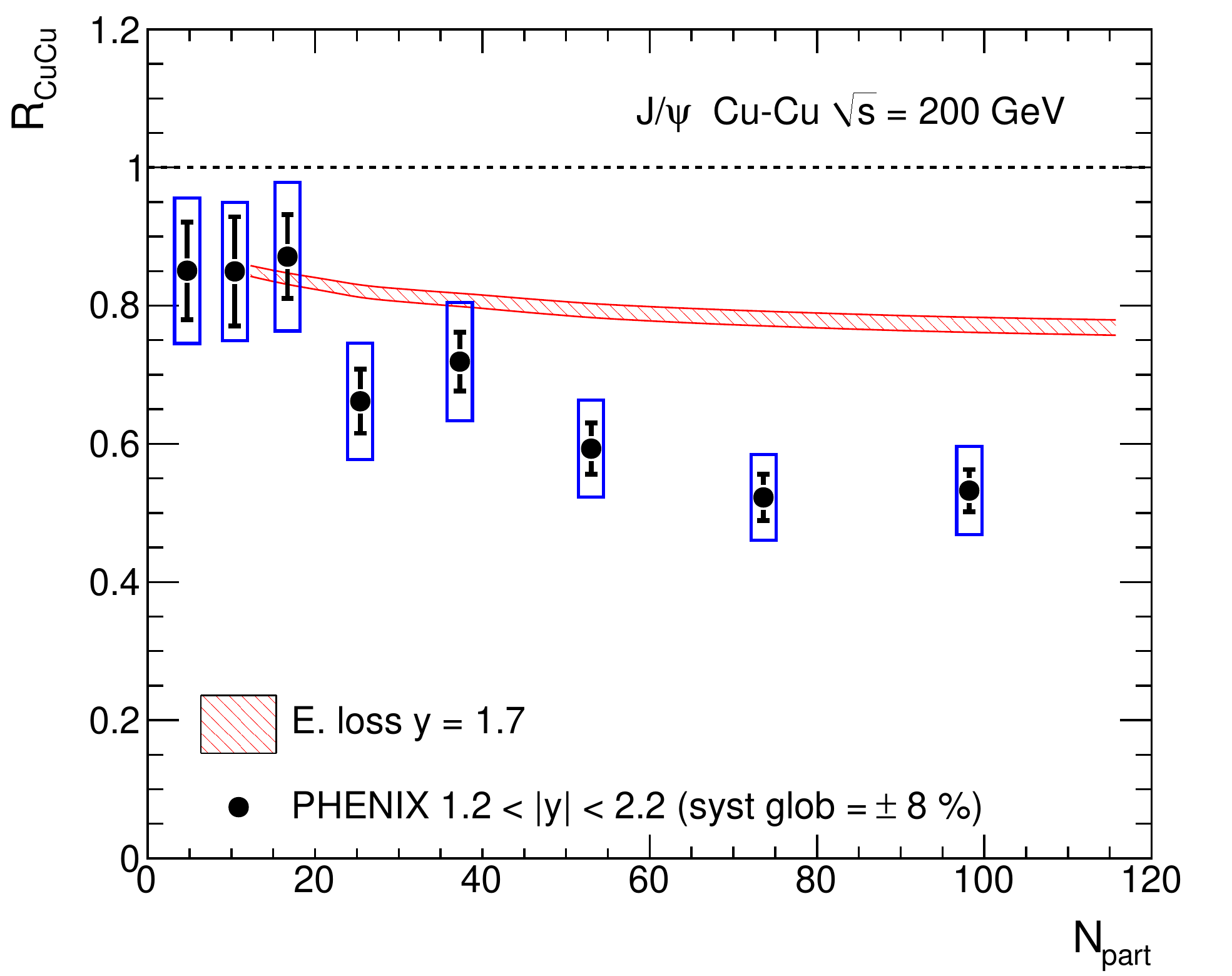}
\includegraphics[width=7.5cm]{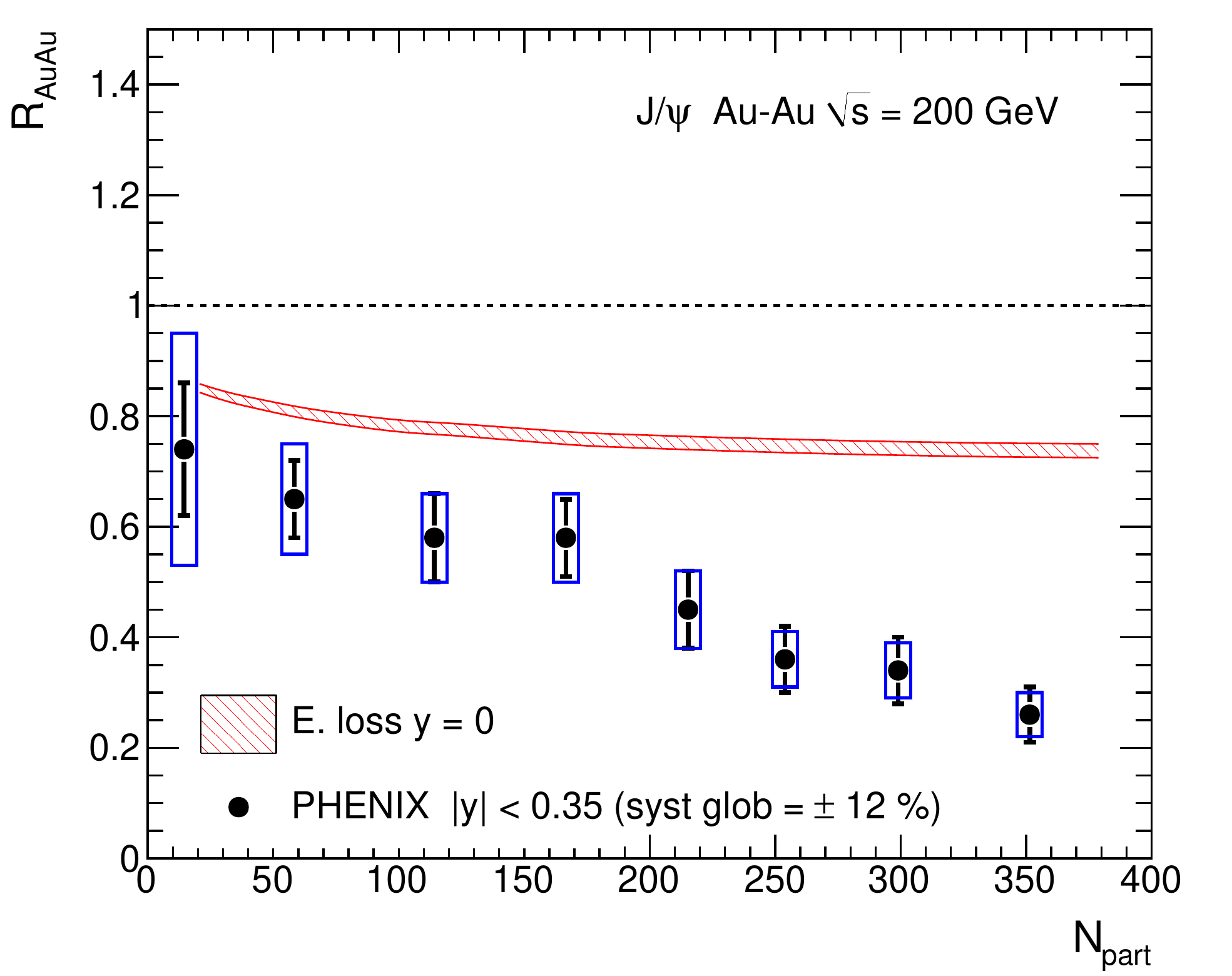}
\includegraphics[width=7.5cm]{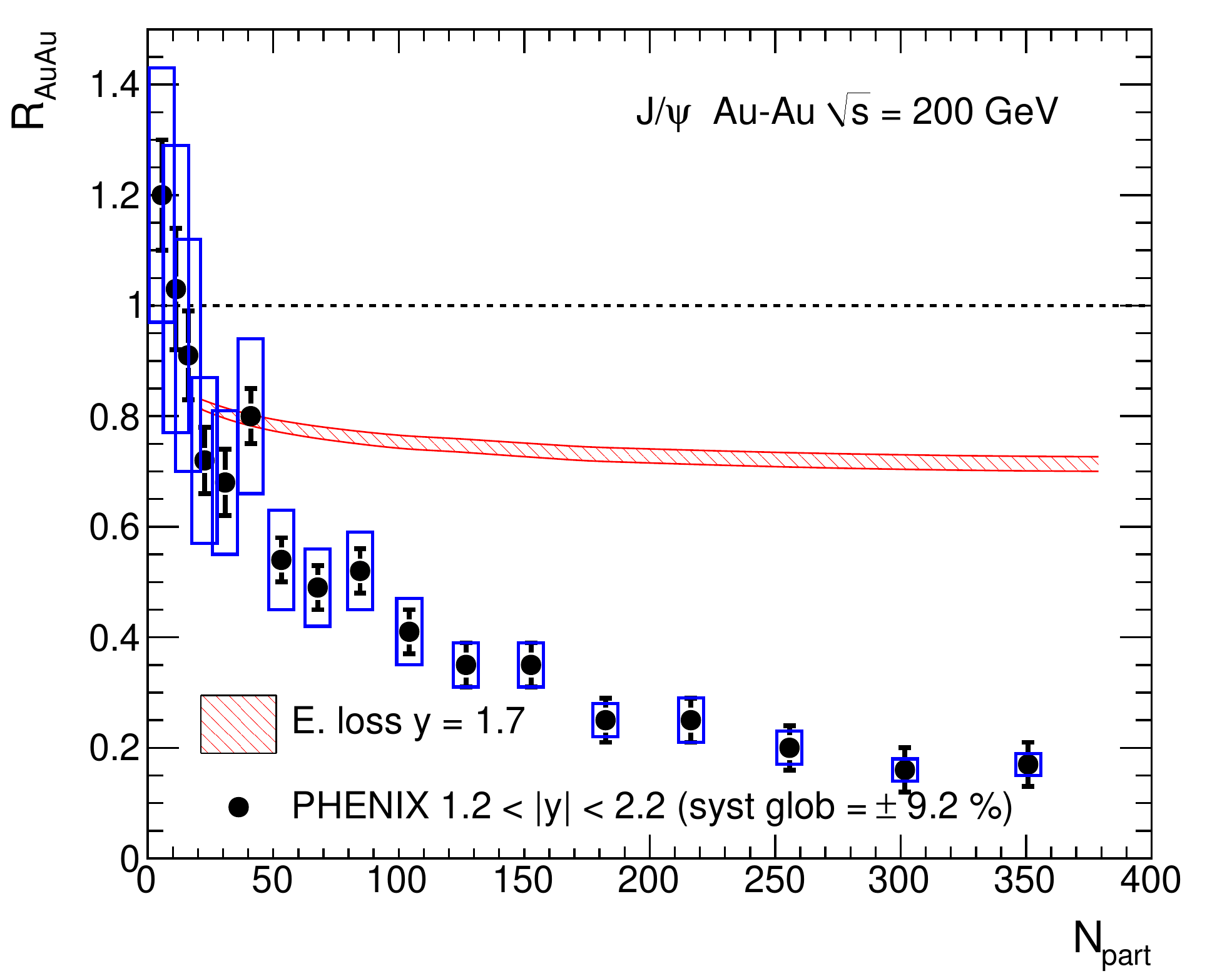}
\includegraphics[width=7.5cm]{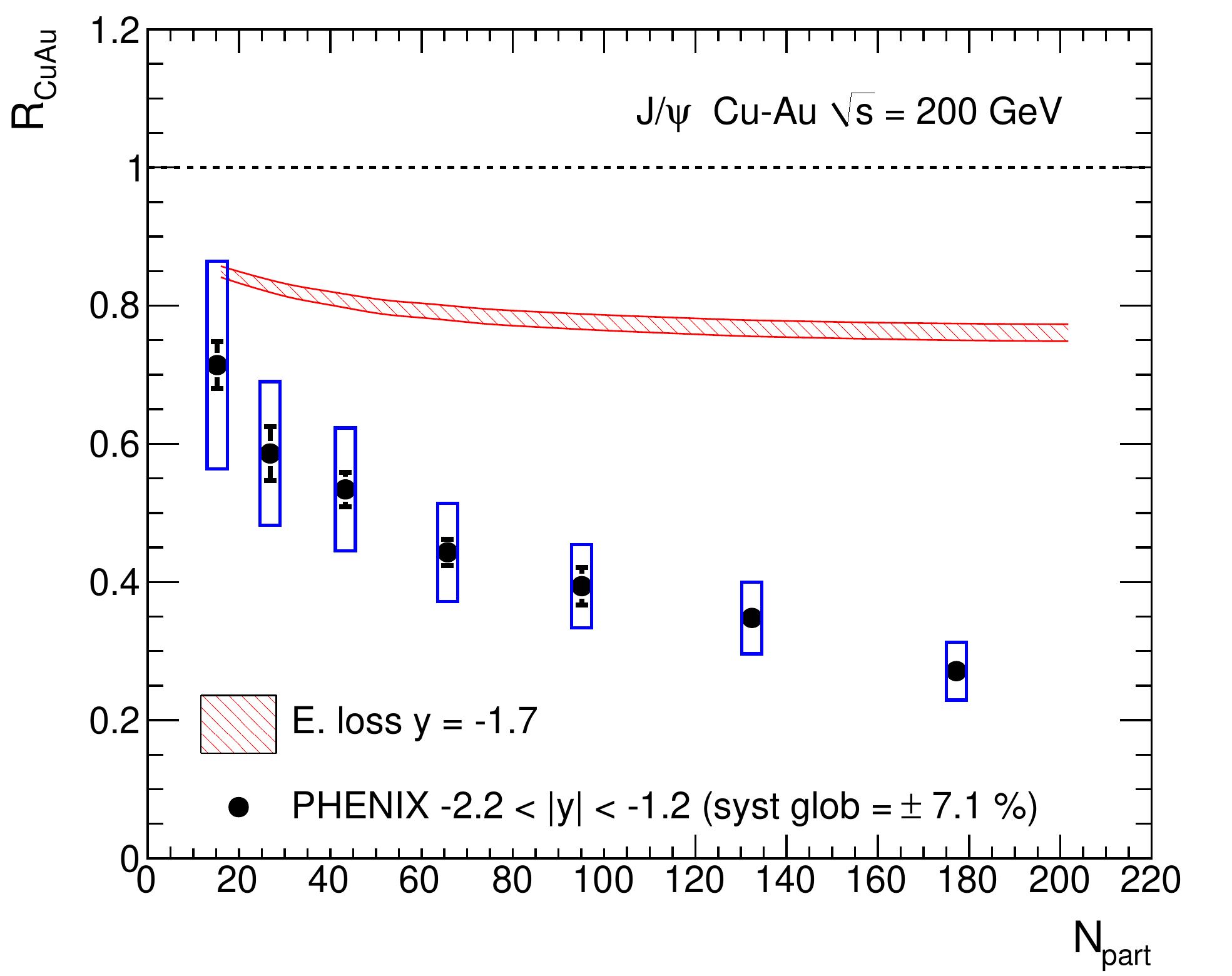}
\includegraphics[width=7.5cm]{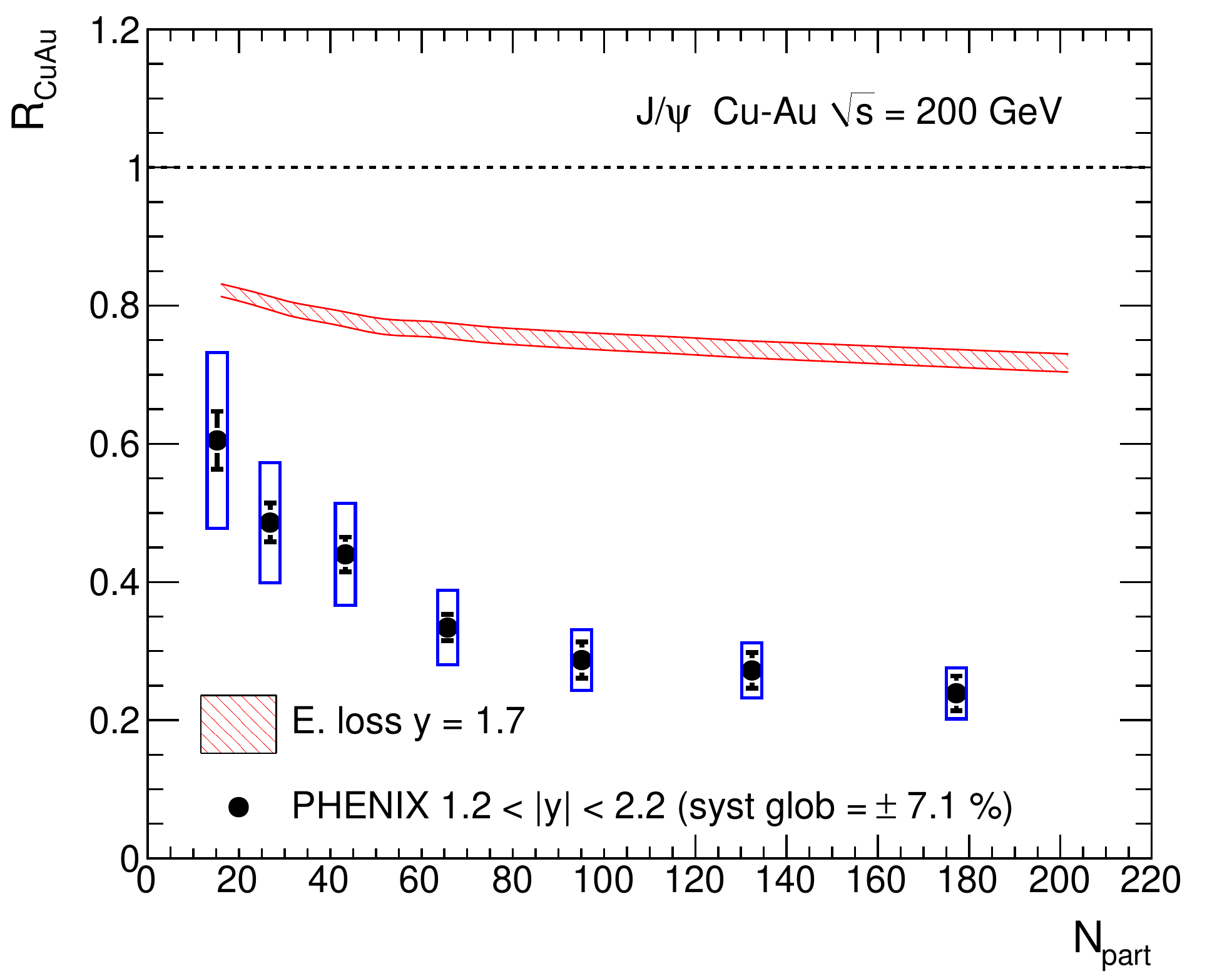}
\caption{Centrality dependence of $\jpsi$ suppression in A--B collisions  ($\sqrt{s}=200$~GeV) predicted in the energy loss model (red band). Top row: Cu--Cu collisions at $y=0$ (left) and $y=1.7$ (right). Middle row:  Au--Au collisions at $y=0$ (left) and $y=1.7$ (right). Bottom row: Cu--Au collisions at $y=-1.7$ (left) and $y=1.7$ (right). PHENIX data are from~\cite{Adare:2008sh,Adare:2006ns,Adare:2011yf,Aidala:2014bqx}.}
\label{fig:rhic_jps_AB_npart}
\end{figure}

The centrality dependence of $\jpsi$ suppression is  computed in Cu--Cu, Au--Au, and Cu--Au collisions, 
see Fig.~\ref{fig:rhic_jps_AB_npart}. 

The $\Npart$ dependence predicted in the model is much less pronounced than the measured one. In Cu--Cu and Au--Au collisions, a discrepancy is seen for $\Npart \gtrsim 60$ while a good agreement is observed for more peripheral collisions. The observed suppression is the strongest in collisions involving the heaviest nucleus (Au). At forward rapidity in Au-Au collisions, the model clearly underpredicts the strength of $\jpsi$ suppression in almost all centrality classes. Despite the fact that the data suffer from a rather large global, systematic uncertainty ranging from $7.1\%$ to $12\%$, the reported $\jpsi$ suppression exceeds significantly the sole effect of parton energy loss in cold nuclear matter.  See section~\ref{sec:discussion} for a discussion. 

\begin{figure}[ht]
\centering
\includegraphics[width=7.5cm]{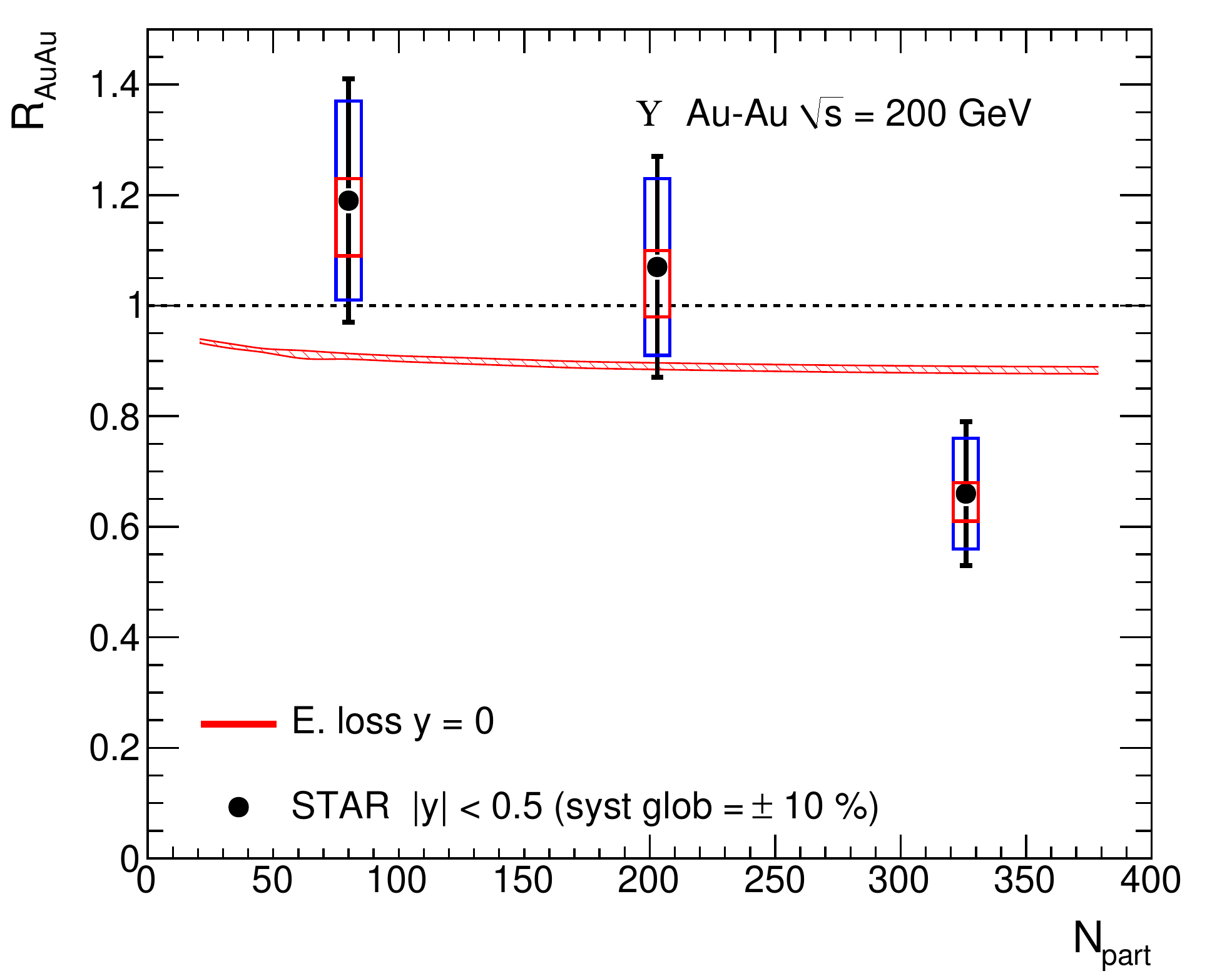}
\caption{Centrality dependence of $\Upsilon$ suppression in Au--Au collisions  ($\sqrt{s}=200$~GeV) predicted in the energy loss model (red band). STAR data are from~\cite{Adamczyk:2013poh}.}
\label{fig:rhic_ups_AB_npart}
\end{figure}

The centrality dependence of $\Upsilon$ suppression is  also computed in Au--Au collisions 
(Fig.~\ref{fig:rhic_ups_AB_npart}) and compared to STAR data~\cite{Adamczyk:2013poh}. In the most central collisions, the reported $\Upsilon$ suppression is stronger than predicted by the energy loss model.

\section{LHC}
\label{sec:lhc}

Predictions for $\jpsi$ and $\Upsilon$ suppression in Pb--Pb collisions ($\sqrts=2.76$~TeV), arising from coherent energy loss, are given as a function of rapidity (section~\ref{sec:lhcrap}) and centrality (section~\ref{sec:lhccentrality}) and compared to ALICE and CMS data.

\subsection{Rapidity dependence}
\label{sec:lhcrap}

\begin{figure}[ht]
\centering
\includegraphics[width=7.5cm]{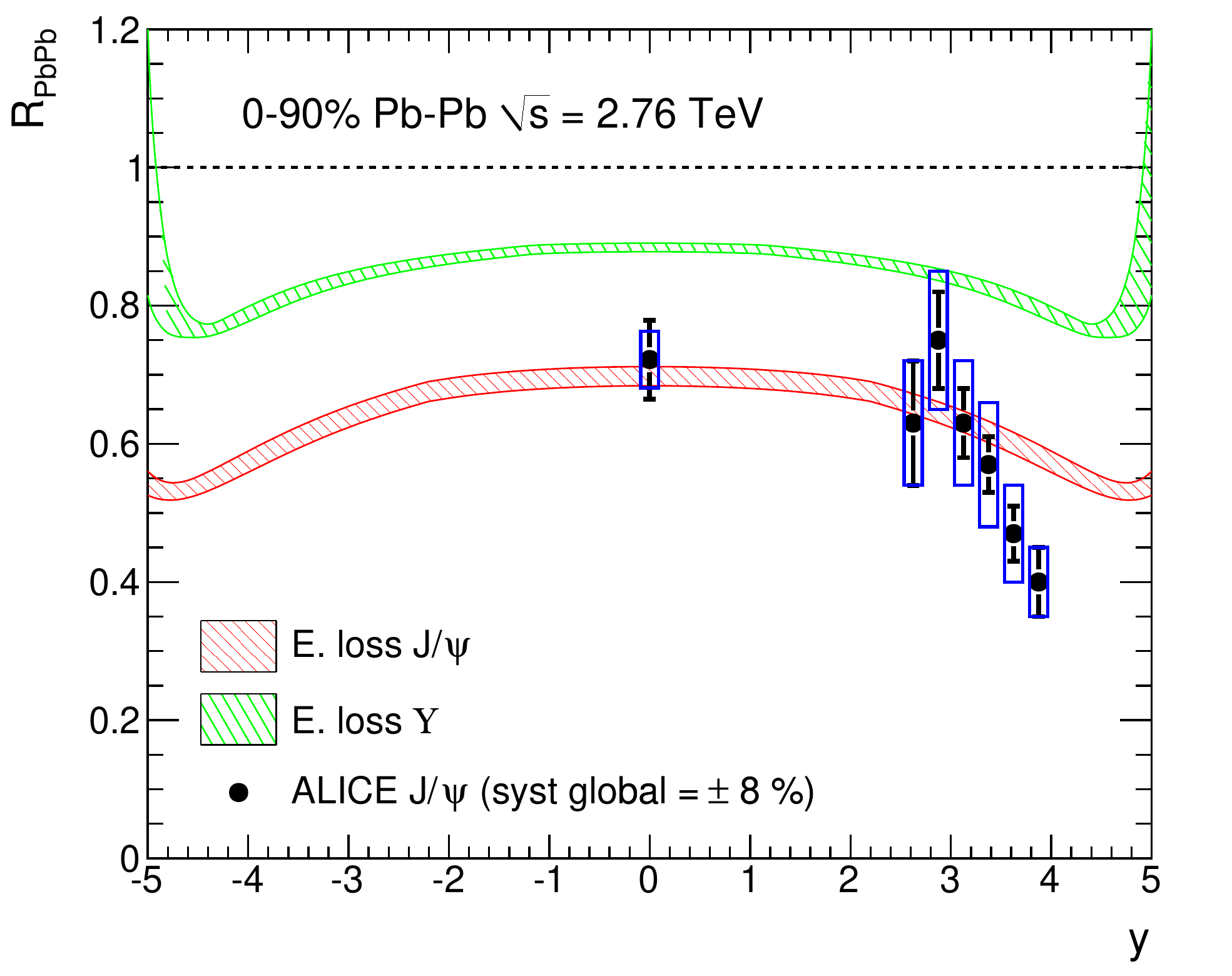}
\caption{Rapidity dependence of $\jpsi$ (lower band) and $\Upsilon$ (upper band) suppression in Pb--Pb collisions at $\sqrt{s}=2.76$~TeV. ALICE $\jpsi$ data are from~\cite{Abelev:2013ila}.}
\label{fig:lhc_jps_ups_pbpb}
\end{figure}

Figure~\ref{fig:lhc_jps_ups_pbpb} shows the rapidity dependence of $\jpsi$ (lower band) and $\Upsilon$ suppression (upper band) expected from energy loss through cold nuclear matter. The different magnitude of the suppression for $\jpsi$ and $\Upsilon$ arises from the mass dependence of coherent energy loss, see Appendix~\ref{app-quenching} and~\cite{Arleo:2012rs}. The quenching factor $\rpbpb$ decreases with rapidity until $|y|\sim5$, from $\rpbpb\simeq 0.7$  down to $\rpbpb\simeq 0.55$ for $\jpsi$ and from 
$\rpbpb\simeq 0.85$ down to $\rpbpb\simeq 0.75$ for $\Upsilon$. 

The $\jpsi$ measurements by ALICE follow the same trend as that of the energy loss model, yet with a stronger rapidity dependence at large $y$. The $\jpsi$ data are well reproduced by the model up to $y\simeq 3$, above which the measured suppression is more pronounced. 

\subsection{Centrality dependence}
\label{sec:lhccentrality}

\begin{figure}[ht]
\centering
\includegraphics[width=7.2cm]{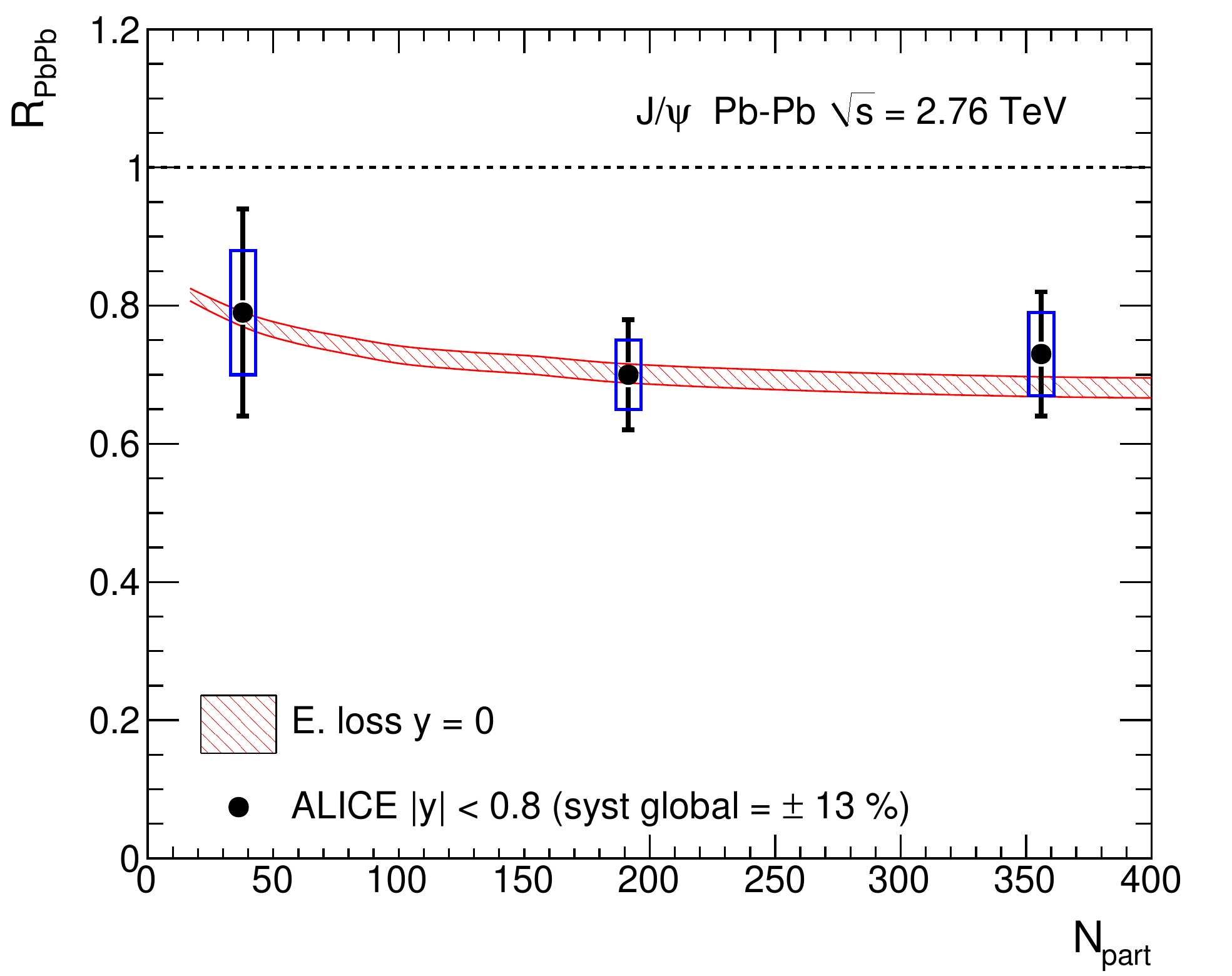}
\includegraphics[width=7.2cm]{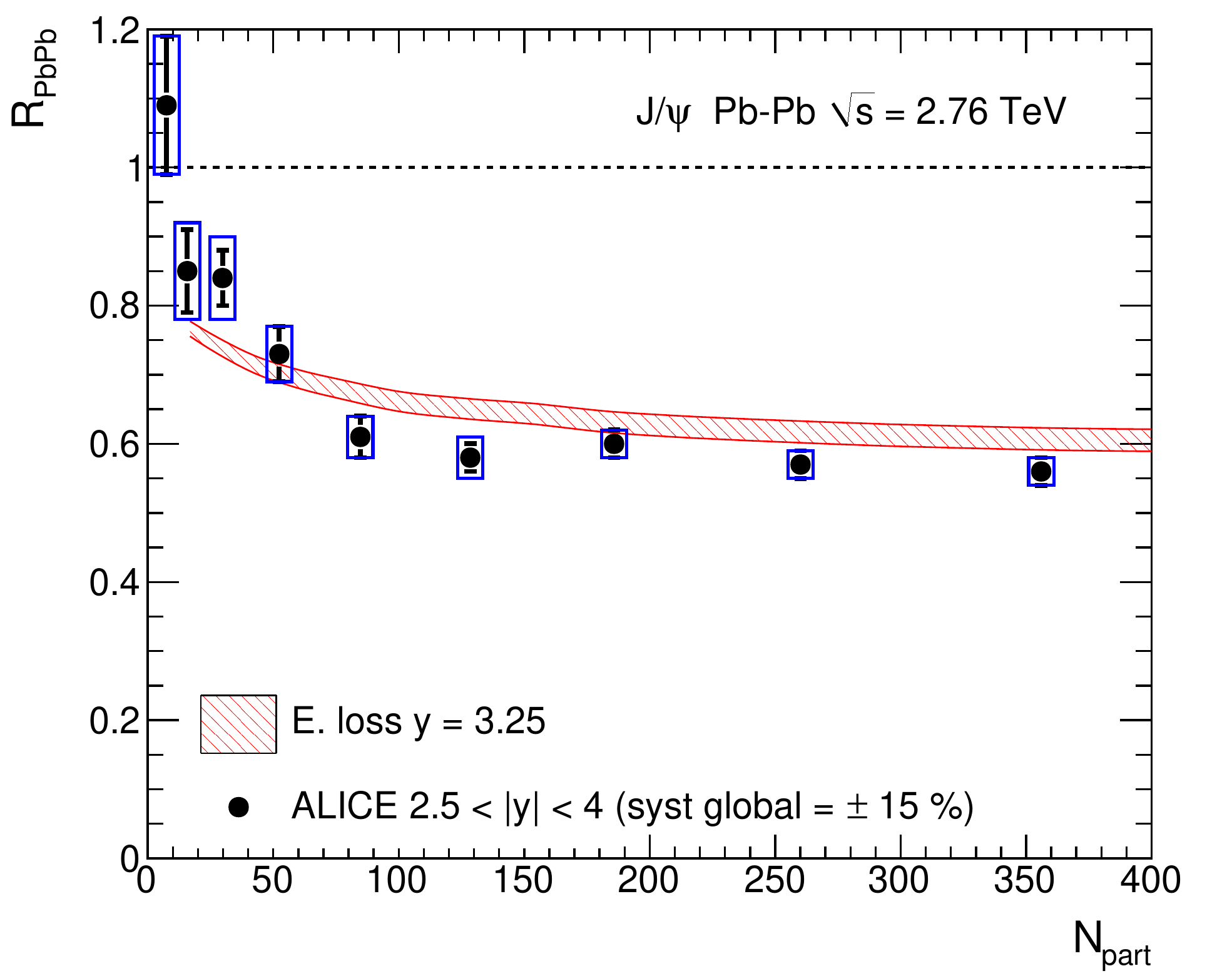}
\caption{Centrality dependence of $\jpsi$ suppression in  Pb--Pb collisions ($\sqrt{s}=2.76$~TeV) at $y=0$ (left) and $y=3.25$ (right) predicted by the energy loss model. ALICE data are from~\cite{Abelev:2013ila}.}
\label{fig:lhc_jps_pbpb_npart}
\end{figure}

The centrality dependence of $\jpsi$ suppression due to energy loss is plotted in Fig.~\ref{fig:lhc_jps_pbpb_npart} at $y=0$ (left) and $y=3.25$ (right) and compared to ALICE measurements~\cite{Abelev:2013ila}. A good agreement is observed in both rapidity bins, although the $\Npart$ dependence at low $\Npart$ is a bit more pronounced in data.

\begin{figure}[h]
\centering
\includegraphics[width=7.5cm]{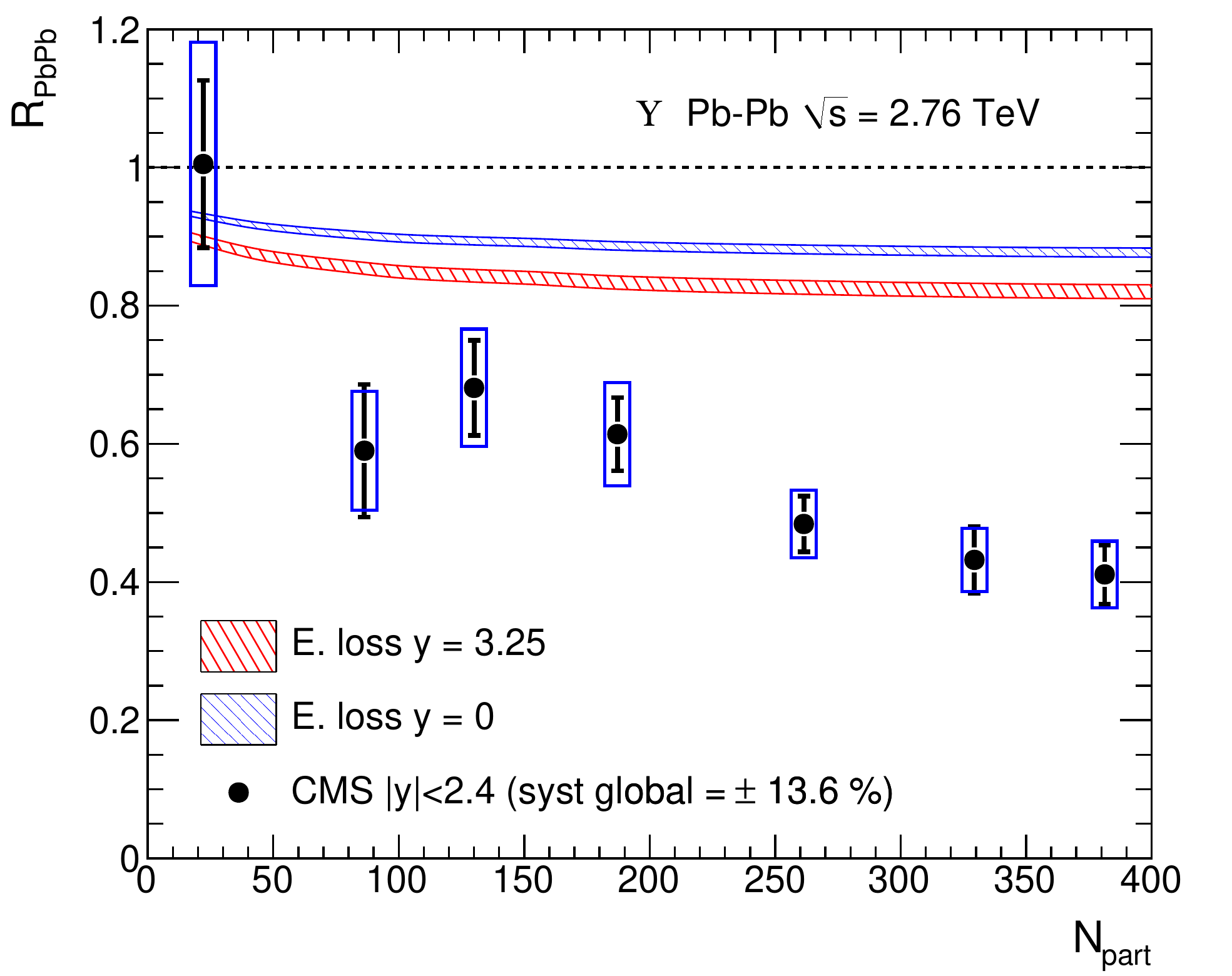}
\caption{Centrality dependence of $\Upsilon$ suppression in  Pb--Pb collisions ($\sqrt{s}=2.76$~TeV) at $y=0$ (upper band) and $y=3.25$ (lower band) predicted by the energy loss model. CMS data are from~\cite{Chatrchyan:2012lxa}.}
\label{fig:lhc_ups_pbpb_npart}
\end{figure} 

The centrality dependence of $\Upsilon$ suppression is computed in Fig.~\ref{fig:lhc_ups_pbpb_npart} at $y=0$ and $y=3.25$ and compared to mid-rapidity ($|y|<2.4$) CMS data~\cite{Chatrchyan:2012lxa}. The suppression reported experimentally is significantly stronger in the data, especially above $\Npart\gtrsim 200$, for which  $\rpbpb\simeq 0.4$--$0.5$ while the model predicts $\rpbpb\simeq 0.9$ at mid-rapidity. This is in sharp contrast with the data--theory comparison in the $\jpsi$ channel.

\section{Discussion}
\label{sec:discussion} 

At RHIC energy, coherent energy loss leads to a sizable $\jpsi$ suppression, with a rather flat dependence both in rapidity (Fig.~\ref{fig:rhic_jps_cucu_auau}) and in centrality (Fig.~\ref{fig:rhic_jps_AB_npart}), for instance $R_{\rm AuAu} \simeq 0.7-0.8$ in a broad centrality and rapidity domain. However, as seen in Figs.~\ref{fig:rhic_jps_cucu_auau} and \ref{fig:rhic_jps_AB_npart}, the predictions of the coherent energy loss model systematically underestimate the strength of $\jpsi$ suppression observed at RHIC (except in the most peripheral collisions), while excellent agreement is reached in d--Au collisions~\cite{Arleo:2012rs}. Here we must recall that the energy loss model applied to A--B collisions is expected to be valid in the restricted region $|y| < |y^{\rm crit}|$, with $y^{\rm crit}  \simeq - 1.2$ at RHIC (see section~\ref{sec:validity}). Since most of the data points lie in the region $|y|\gtrsim |y^{\rm crit}|$, $\jpsi$ hadronization might occur inside one of the two nuclei leading to some additional suppression coming from nuclear absorption. It seems however unlikely that nuclear absorption effects could fill the gap between the energy loss model predictions and 
Cu--Au or Au--Au data for $\jpsi$ suppression in this rapidity domain. 

We conclude that the large difference between the energy loss model predictions and the data is qualitatively consistent with the onset of {\it hot} effects such as Debye screening or gluon dissociation in the hot medium formed in heavy-ion collisions at RHIC. In this respect, it is noticeable in Fig.~\ref{fig:rhic_jps_AB_npart} that the energy loss expectations are consistent with the data in peripheral collisions, but strongly deviate from them in more central collisions.   

\begin{figure}[h]
\centering
\includegraphics[width=7.3cm]{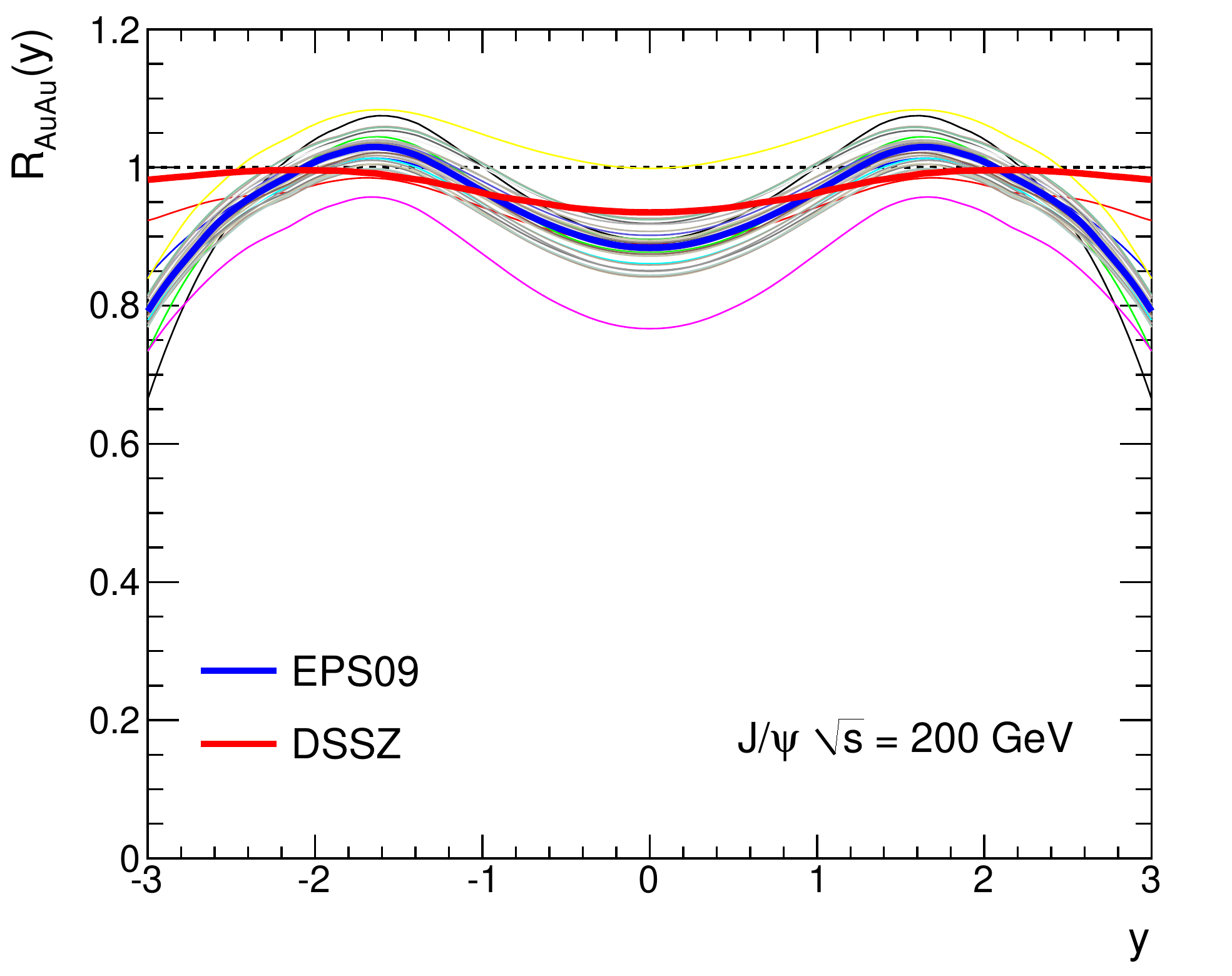} \hskip 3mm \includegraphics[width=7.3cm]{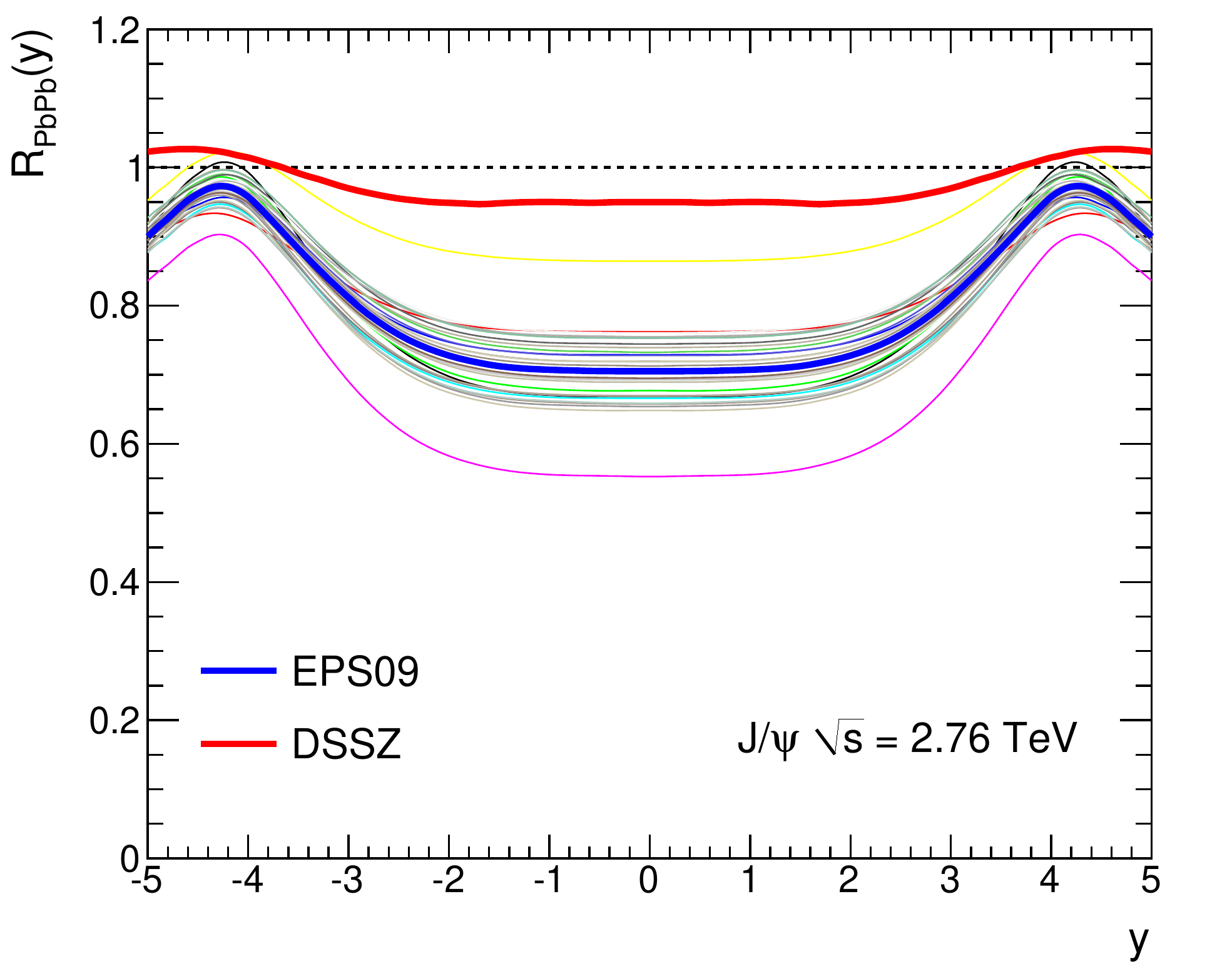} 
\caption{Rapidity dependence of $\jpsi$ suppression in minimum bias Au--Au collisions at $\sqrt{s}=200$~GeV (left) and Pb--Pb collisions at $\sqrt{s}=2.76$~TeV (right) estimated 
from the sole effect of nPDFs. The EPS09 and DSSZ central sets are shown as thick solid lines and EPS09 error sets as thin solid lines.}
\label{fig:npdf-effects}
\end{figure}

At the LHC, a relatively smooth dependence of quarkonium suppression in $y$ or centra\-li\-ty is also predicted in the energy loss model, see Figs.~\ref{fig:lhc_jps_ups_pbpb}, \ref{fig:lhc_jps_pbpb_npart} and \ref{fig:lhc_ups_pbpb_npart}. However, the dependence of 
$R_{\rm AA}$ 
on rapidity is less flat than at RHIC, with a decrease up to the largest $|y| \sim 3-4$ where the model can be applied ($|y^{\rm crit}| \simeq 3.8$ at $\sqrt{s}=2.76$~TeV, see section~\ref{sec:validity}). This is at variance with calculations based on nPDF effects which predict a rise with $|y|$, see for instance~\cite{Abelev:2012rv} and Fig.~\ref{fig:npdf-effects} (right) below. We stress that at LHC the sole effect of energy loss is responsible for a large $\jpsi$ suppression, $R_{\rm PbPb} \simeq 0.6-0.7$  in a broad centrality and rapidity domain. 

Quite surprisingly, there is a very good agreement between the ALICE data and the energy loss model for $\jpsi$ suppression in the domain $|y|\lesssim 3$, see Figs.~\ref{fig:lhc_jps_ups_pbpb} and \ref{fig:lhc_jps_pbpb_npart}. This indicates no {\it net} hot medium effect in the $\jpsi$ channel. In other words, if large hot medium effects are at work in $\jpsi$ production at $|y| \lesssim 3$, they apparently roughly compensate one another. For instance, the expected $\jpsi$ suppression from dissociation processes or screening effects in the hot medium might be compensated by the recombination of charm quark pairs. Thus, the energy loss model predictions for $\jpsi$ suppression in A-B collisions are consistent with the presence of recombination at LHC, such a large recombination effect being not required at RHIC. 

In the $\Upsilon$ channel, the suppression measured by CMS is much stronger than the energy loss model prediction, Fig.~\ref{fig:lhc_ups_pbpb_npart}.  This may be qualitatively understood by the negligible recombination rate of bottom quark pairs when compared to charm quark pairs, leading to some sizable {\it net} hot medium effect in the $\Upsilon$ channel.

In the present study we focussed 
on the effect of energy loss through cold nuclear matter. Of course, the magnitude of the obtained suppression could possibly be affected by additional nPDF effects. 
We estimated simply
the amount of $\jpsi$ suppression resulting from nPDF effects 
alone\footnote{In this illustration, nPDF effects are given by $R_{\rm g}^{\rm Pb}(x_1, \mT^2)\times R_{\rm g}^{\rm Pb}(x_2, \mT^2)$ with the momentum fractions $x_{1, 2} = \mT/\sqrt{s}\times \exp\left(\pm y\right)$.} 
using two recent next-to-leading order sets of nuclear parton densities, EPS09~\cite{Eskola:2009uj} and DSSZ~\cite{deFlorian:2011fp}. (For more calculations involving nPDFs, the reader may refer to~\cite{Vogt:2010aa}.)
The central EPS09 and DSSZ predictions are shown in Fig.~\ref{fig:npdf-effects} as thick solid lines. In addition, the calculation has been carried out using the 30 (equally likely) EPS09  error sets coming from the (positive and negative) variation of the 15 parameters used in the global fit analysis.\footnote{The variation of each parameter in the EPS09 global fit is such that the $\chi^2$ function increases by $\Delta \chi^2 = 50$ from its minimum, corresponding to a 90\% confidence criterion~\cite{Eskola:2009uj}.} As can be seen in Fig.~\ref{fig:npdf-effects}, the bulk of nPDF calculations points to a moderate 5--15\% $\jpsi$ suppression at RHIC. At the LHC, using the DSSZ 
set leads to less than 5\% suppression, while all but two of the (equally likely) EPS09 error sets predict $\rpbpb \simeq 0.65$--0.85 at $|y|\leq 3$. 

At RHIC and the LHC, the expected 
magnitude of the suppression due to nPDF effects alone never exceeds (and is actually most often {\it smaller} than) that due to energy loss, compare for instance Fig.~\ref{fig:rhic_jps_cucu_auau} to Fig.~\ref{fig:npdf-effects} (left) and Fig.~\ref{fig:lhc_jps_ups_pbpb} (red band) to Fig.~\ref{fig:npdf-effects} (right). At RHIC, energy loss 
is likely to be more important than nPDF effects. At the LHC, the 
energy loss effect is likely to be as important (using EPS09) or more important (using DSSZ) than nPDF effects at mid-rapidity, and more important (for any nPDF choice) at large enough $|y|$. We thus believe our above qualitative discussion to be unaffected by the inclusion of nPDF effects. Our study emphasizes 
that coherent energy loss effects should be taken into account in order to obtain a reliable baseline for quarkonium suppression in heavy-ion collisions.

Finally, apart from the fact that energy loss is likely to be quantitatively 
as  important (or more) than nPDF effects, we 
would like to stress that the combined effect of energy loss + nPDF should \emph{not} be estimated by multiplying the two associated $R_{\rm AA}$. Indeed, see the case of p--A collisions considered in \cite{Arleo:2012rs}. It was found there that for consistency, the predictions in a model ``energy loss + nPDF" should be made with a different (smaller) value of $\qhat_0$, obtained from a fit of the ``energy loss + nPDF" model to E866 data
for $\jpsi$ suppression in proton--tungsten collisions. 
The output is that the energy loss predictions for the suppression with or without nPDF effects were of similar magnitude at RHIC and LHC. We expect a similar effect in A--B collisions. 

\acknowledgments

This work is funded by ``Agence Nationale de la Recherche'' (grant ANR-PARTONPROP). 
\appendix

\section{Quenching weight}
\label{app-quenching}

The quenching weight ${\cal P}(\varepsilon, E, \ell^2)$ is related to the radiation spectrum ${\dd I}/{\dd \varepsilon}$ as~\cite{Arleo:2012rs}
\be
\label{relation-P-spec}
{\cal P}(\varepsilon, E, \ell^2) = \frac{\dd I}{\dd\varepsilon} \, \exp \left\{ - \int_{\varepsilon}^{\infty} \dd\omega  \frac{\dd{I}}{\dd\omega} \right\} = \frac{\partial}{\partial \varepsilon} \, \exp \left\{ - \int_{\varepsilon}^{\infty} \dd\omega  \frac{\dd{I}}{\dd\omega} \right\} \, .
\ee 
The medium-induced, coherent radiation spectrum reads\footnote{Note that if another hard colored particle is produced in association with the $Q\bar{Q}$ pair (for instance for $\psi$ produced at large $\pt$) the medium-induced gluon spectrum is similar to~\eq{spectrum-def} yet with a different color and kinematical prefactor~\cite{Liou:2014rha,Peigne:2014rka}.}
\be
\label{spectrum-def}
\frac{\dd I}{\dd \varepsilon} = \frac{N_c \alpha_s}{\pi \varepsilon} \left\{ \ln{\left(1+\frac{\ell^2 E^2}{M_\perp^2 \varepsilon^2}\right)} - \ln{\left(1+\frac{\Lambda_{\rm p}^2 E^2}{M_\perp^2 \varepsilon^2}\right)} \right\} \, \Theta(\ell^2 - \Lambda_{\rm p}^2) \, ,
\ee
where $\ell$ denotes the transverse momentum broadening acquired when crossing the target nucleus ($\ell = \ell_{\rm B}$ for target nucleus B), and $\Lambda_{\mathrm p}^2 = {\rm max}(\Lambda_{_\mathrm{QCD}}^2,\ell_{\rm p}^2)$~\cite{Arleo:2012rs}. The medium-induced spectrum \eq{spectrum-def} is defined for a target nucleus B with respect to a target proton, and vanishes when B$=$p.  

We easily check that $E\,{\cal P}(\varepsilon,E,\ell^2)$ is a scaling function of $x \equiv \varepsilon/E$ and $\ell^2$, and introduce the function $\Phat$,
\be
\label{Phat-def}
\Phat(x,\ell^2) \equiv E\,{\cal P}(\varepsilon,E,\ell^2) \, .
\ee
Using \eq{relation-P-spec}, \eq{spectrum-def} and \eq{Phat-def}, we get the explicit form of the `quenching weight' $\Phat(x,\ell^2)$ to be used in our study,
\bea
\label{Phat-explicit}
\Phat(x,\ell^2) &=& \frac{N_c \alpha_s}{\pi  x} \ln{\left( \frac{x^2 M_\perp^2 + \ell^2}{x^2 M_\perp^2 + \Lambda_{\rm p}^2}\right)} \exp \left\{ - \int_{x}^{\infty} \dd v \, \frac{N_c \alpha_s}{\pi v} \ln{\left( \frac{v^2 M_\perp^2 + \ell^2}{v^2 M_\perp^2 + \Lambda_{\rm p}^2}\right)} \right\} \nn \\
&=&  \frac{\partial}{\partial x} \, \exp \left\{ - \frac{N_c \alpha_s}{2\pi} \int_{\Lambda_{\rm p}^2/(x^2 M_\perp^2)}^{\ell^2/(x^2 M_\perp^2)} \frac{\dd t}{t} \ln(1+t)  \right\} \, ,
\eea
where the $\Theta$-function appearing in \eq{spectrum-def} is now implicit. 
$\Phat$ can be expressed in terms of the dilogarithm ${\rm Li}_2(x) = - \int_0^x \frac{\dd t}{t} \ln(1-t)$ as
\be
\label{quenching-dilog}
\Phat(x,\ell^2) =  \frac{\partial}{\partial x} \, \exp \left\{\frac{N_c \alpha_s}{2 \pi} \left[ {\rm Li}_2\left(\frac{-\ell^2}{x^2 M_\perp^2} \right) - {\rm Li}_2\left(\frac{-\Lambda_{\rm p}^2}{x^2 M_\perp^2} \right) \right] \right\} \, .
\ee
Finally, let us mention that when the transverse momentum broadening $\ell$ becomes {\it very large} (\ie, for a very large target nucleus), and $x = \varepsilon/E$ satisfies $\Lambda_{\rm p}^2/M_\perp^2 \ll x^2 \ll \ell^2/M_\perp^2$, the quenching weight can be approximated by the simple expression (use \eq{Phat-explicit}) 
\be
\label{Phat-approx}
\Phat(x,\ell^2) \simeq \frac{\partial}{\partial x} \, \exp \left\{ - \frac{N_c \alpha_s}{\pi}  \ln^2{\left( \frac{\ell}{x M_\perp}\right)}  \right\} \, .
\ee
For realistic nuclear sizes however, we have $\Lambda_{\rm p}^2 \lsim \ell^2$ rather than the strong inequality $\Lambda_{\rm p}^2 \ll \ell^2$, and the exact expression \eq{quenching-dilog} should be preferred to the approximation \eq{Phat-approx}. 

\providecommand{\href}[2]{#2}\begingroup\raggedright\endgroup

\end{document}